\documentclass[12pt,epsf]{article}

\usepackage{epsfig}

\def\ss{\scriptscriptstyle}

\def\thm{\theta_\mu}
\def\tha{\theta_A}
\def\gev{{\rm \, Ge\kern-0.125em V}}
\def\tev{{\rm \, Te\kern-0.125em V}}
\def\ga{\mathrel{\raise.3ex\hbox{$>$\kern-.75em\lower1ex\hbox{$\sim$}}}}
\def\la{\mathrel{\raise.3ex\hbox{$<$\kern-.75em\lower1ex\hbox{$\sim$}}}}

\def\NPB{{\em Nucl.\ Phys.} B}
\def\PLB{{\em Phys.\ Lett.} B}

\def\PRD{{\em Phys.\ Rev.} D}

\def\PR{{\em Phys.\ Rep.}}

\def\Yi{\eta^{\ast}_{11} \left( \frac{y_{i}}{2} g' Z_{\chi 1} + 
        g T_{3i} Z_{\chi 2} \right) + \eta^{\ast}_{12} 
        \frac{g m_{q_{i}} Z_{\chi 5-i}}{2 m_{w} B_{i}}}

\def\Xi{\eta^{\ast}_{11} 
        \frac{g m_{q_{i}}Z_{\chi 5-i}^{\ast}}{2 m_{w} B_{i}} - 
        \eta_{12}^{\ast} e_{i} g' Z_{\chi 1}^{\ast}}

\def\Wi{\eta_{21}^{\ast}
        \frac{g m_{q_{i}}Z_{\chi 5-i}^{\ast}}{2 m_{w} B_{i}} -
        \eta_{22}^{\ast} e_{i} g' Z_{\chi 1}^{\ast}}

\def\Vi{\eta_{22}^{\ast} \frac{g m_{q_{i}} Z_{\chi 5-i}}{2 m_{w} B_{i}}
        + \eta_{21}^{\ast}\left( \frac{y_{i}}{2} g' Z_{\chi 1}
        + g T_{3i} Z_{\chi 2} \right)}

\def\zthree{\delta_{1i} [g Z_{\chi 2} - g' Z_{\chi 1}]}

\def\zfour{\delta_{2i} [g Z_{\chi 2} - g' Z_{\chi 1}]}

\begin{document}
\begin{titlepage} 
\pagestyle{empty}
\rightline{UMN--TH--1811/99}
\rightline{TPI--MINN--99/36}
\rightline{MADPH-99-1125}
\rightline{hep-ph/9908311}
\rightline{August 1999}  

\vspace{1.5cm}
\begin{center}
\baselineskip=21pt
{\large {\bf Variations of the Neutralino 
Elastic Cross-Section with CP Violating Phases}}\\
\vspace{1cm}
{Toby Falk}\\
{\it Department of Physics, University of Wisconsin, Madison, WI 53706, USA}\\ 
\vspace{.5cm}
{Andrew Ferstl} and {Keith A. Olive}\\
{\it Theoretical Physics Institute, School of Physics and Astronomy, \\
University of Minnesota, Minneapolis, MN 55455, USA}\\ 
\vspace*{.5in}
{\bf Abstract}
\end{center}

\baselineskip=18pt
\noindent
We analyze the neutralino-nucleus elastic cross-section
in the MSSM, including contributions from CP-violating phases, using
the four-fermi neutralino-quark interaction.  Over a wide range of the MSSM 
parameter space we show the variations in the  cross-sections due
to these phases.  We further concentrate on the regions which are consistent
with constraints from the electric dipole moment of the electron, neutron and
mercury atom. In the regions we examine in detail, we find suppressions by up to a factor of two,
while enhancements in the cross-sections are no greater than
$\sim$ 10\%.
\vfill
\end{titlepage}

\normalsize
\section{Introduction}
\baselineskip=18pt

Supersymmetry provides a framework for 
constructing new models for physics beyond the standard model.  The
motivations for supersymmetry are well known, and include a stabilization of 
the mass hierarchy problem and unification of gauge couplings at the
GUT scale.  In the minimal supersymmetric extension of the standard model
(MSSM) \cite{MSSM},  conservation of R-parity, which can be related to baryon number,
lepton number and spin, guarantees that the  lightest supersymmetric partner
(LSP) is stable and  provides a candidate for the 
non-baryonic dark matter in the universe.  Neutralinos, mixtures of neutral gauginos and
Higgsinos, are the best motivated LSP candidate in the MSSM
\cite{ehnos}.

 For a given set of MSSM parameters, the exact identity of the LSP
is determined, its annihilation cross-section and hence its relic density
can be calculated, as can its interactions with ordinary matter.  The latter
is key in efforts to detect MSSM dark matter in laboratory experiments
\cite{suppl,review}. Indeed, over the years a great effort has been made to
study the elastic cross-section of the LSP with nuclei in order to predict
the possible detection rates in cryogenic detectors. Recently, it has been
shown \cite{FFO,CIN,KS}
that the CP  violating phases in the MSSM may have significant 
effects on the detection rate of the neutralino.  
Here we continue our investigation of the neutralino-nucleon elastic
cross-section over a wide range of MSSM parameters and show  that these
phases can give significant,  variations to the cross-section. We also
take into account the  experimental electric dipole constraints of the
electron, neutron and mercury atom, and show that variations in the
cross-sections of order 0.4 -- 1.1 persist.  

CP violating phases can arise from several sources in the MSSM:  the Higgs
mixing mass $(\mu)$  in the superpotential, and the supersymmetry
breaking parameters which includes the gaugino masses ($M_{i}$),
the soft scalar masses, the bilinear term ($B\mu$), and
the trilinear terms ($A_{f}$).  Not all of these phases are physical \cite{dgh}.
For our purposes, the phases
associated with the gaugino masses
are rotated away and $B\mu$ will be taken as real 
so that the vacuum expectation values of the Higgs fields
are real.  
Also, the phases of the trilinear terms, $A_{f}$ (with the exception of $A_t$),
will be set equal to one another, for simplicity.  The trilinear stop mixing
parameter
$A_t$ has a (real) quasi-fixed point in its evolution from the GUT scale to
the electroweak scale, so we take $A_t$ real. 
 Then, the only physical phases in the MSSM are the phase 
associated with $\mu$ ($\theta_{\mu}$) and the phase associated
with the $A_{f}$ ($\theta_{A}$).   
Typically, in order to satisfy 
constraints\cite{smallthmu} on the electric dipole moment (EDM) of
the electron, neutron and $^{199}$Hg atom,
CP violating phases
must take on very small values (on the 
order of $ 10^{-2}-10^{-3}$ for MSSM parameters on the order of 
100 GeV). This limit can be weakened \cite{nath,KO} if
the SUSY masses are taken to be upwards of a TeV, which is not a
cosmological viability if the neutralino is gaugino-like\cite{FOS}. 
Models with heavy first two generations of sfermions and a
light third generation\cite{heavy}, on the other hand, may allow large
phases while permitting the neutralino to annihilate into e.g. tau
pairs, so that the
cosmological constraints are satisfied. Alternatively, with two CP
violating phases and several contributions  to the electric dipole
moments cancellations can occur and
 allow the CP violating
phases to take on larger values \cite{fko1} - \cite{prs}.
We neglect the (typically small)
phase misalignment between the Higgs vevs, which is induced by one loop
corrections
to the Higgs potential in the presence of CP violation in the stop
sector \cite{phm,pw}.  The effect
of this misalignment may be significant for large $|\mu A_t|/m_A^2$,
particularly at large $\tan\beta$. 

Previous studies \cite{FFO,CIN} have shown that the CP-violating
phases can have a large effect on neutralino-nucleus elastic
scattering cross-sections. In certain regions of the parameter space,
the elastic cross-section with matter can show a dramatic dependence
(usually a minimum) as a function of the CP violating phase $\thm$.  Here,
this effect is studied in more detail. In particular, we determine the
dependence of the cross-section on the phases across the $M_2$-$\mu$ plane
in the MSSM, including a separate investigation of the phase
dependence of the 
spin-dependent and spin-independent cross-sections. We also examine
the effect of the constraints of the electric dipole moments on the
variation of the cross-section with respect to the phases.
Generically,  the EDM constraints require the phases to be
sufficiently small that the cross-sections are only very mildly
affected.  Nevertheless, we do find
parameter  regions, predominantly where the neutralino is
gaugino(bino)-like, where the constraints on the CP violating phases
can be satisfied, and where the phases still produce significant
changes to the scattering cross-section. In these regions, we find that the
elastic cross-section may be enhanced by  $\sim$ 10\% or reduced by
as much as a factor of 2.5.       

In Section 2, the effective neutralino-quark interaction Lagrangian
is presented and the dominant terms in the direct detection rate are
discussed.
In Section 3, the analytic expression 
for the neutralino-nucleus scattering cross-section is found using the
low  energy effective Lagrangian. Finally, in Section 4 
the variation of the elastic cross-section with the CP
violating phases is shown. 
First, the effect of the phases
on the spin-dependent, spin-independent, and total scattering rates 
is displayed for a particular set of parameters on the $M_{2} - \mu$
plane.  For these plots, no EDM constraints are imposed so that the
general behavior with respect to the CP violating phases can be shown.  
Then a general scan of parameter space is performed
while imposing the electron, neutron and mercury EDM constraints so
that a consistent and realistic understanding of the effect of 
CP-violation can be shown. 

\section{Effective Lagrangian}

The neutralino LSP is the
lowest mass eigenstate of the linear 
combination of the bino, wino, and Higgsinos.  In
our notation,
 \begin{equation}
\chi = Z_{\chi 1}\tilde{B} + Z_{\chi 2}\tilde{W} +
Z_{\chi 3}\tilde{H_{1}} + Z_{\chi 4}\tilde{H_{2}}
\label{id}
\end{equation}
The neutralino mass matrix in the $(\tilde{B}, \tilde{W},
\tilde{H_{1}}, \tilde{H_{2}})$ basis is:

\[ N = \left( \begin{array}{cccc}
M_{1} & 0 & -M_{Z} \sin{\theta_{W}} \cos{\beta} & 
M_{Z} \sin{\theta_{W}} \sin{\beta} \\
0 & M_{2} & M_{Z} \cos{\theta_{W}} \cos{\beta} &
-M_{Z} \cos{\theta_{W}} \sin{\beta} \\
-M_{Z} \sin{\theta_{W}} \cos{\beta} & 
M_{Z} \cos{\theta_{W}} \cos{\beta} & 0 & - \mu \\
M_{Z} \sin{\theta_{W}} \sin{\beta} &
-M_{Z} \cos{\theta_{W}} \sin{\beta} & - \mu & 0
\label{neutralinomass}
\end{array} \right), \] 
where $M_{1}$ and $M_{2}$ are the U(1) and SU(2) gaugino masses,
$\tan{\beta}$ is the ratio of the Higgs vacuum expectation values, and
$\mu$ is the Higgsino mixing mass.  We assume a common gaugino mass at
the GUT scale, which gives $M_1 = \frac{5}{3}\tan^2{\theta_{W}}M_{2}$, and
we take $\tan\beta, \mu$, and $M_2$ as free parameters.  The mass matrix is
diagonalized by a matrix $Z$, $Z^* N Z^{-1}$, which determines the eigenstate
composition in (\ref{id}).

The sfermion mass$^{2} $ matrix can be written as
\[ M^2 = \smallskip\pmatrix{ M_L^2 + m_f^2 + \cos 2\beta (T_{3f} - Q_f\sin ^2
\theta_W) M_Z^2 & -m_f\,\overline{m}_{\ss f} e^{i \gamma_f}
\cr
\noalign{\medskip} -m_f\,\overline{m}_{\ss f} e^{-i \gamma_f} & M_R^2 + m_f^2 +
Q_f\sin ^2
\theta_W M_Z^2 \cos 2\beta 
\cr }.
\label{sfermionmass}\]
Here, $ M_{L(R)} $ are the soft supersymmetry breaking sfermion masses which 
are real since we have assumed that they are generation independent and 
generation diagonal.  Also,
\begin{equation}
\overline{m}_{f} e^{i \gamma_{f}} = R_{f} \mu + A_{f}^{\ast} = 
R_{f} \left| \mu \right| e^{i \theta_{\mu}} + 
\left| A_{f} \right| e^{-i \theta_{A}}
\label{offdiag}
\end{equation}
where $ m_{f} $ is the mass of the fermion and $ R_{f} = \cot{\beta}
(\tan{\beta})$ for weak isospin +1/2 (-1/2) fermions.  For simplicity,
we will choose all $A_{f}$, except
for $A_{t}$, to be degenerate at the weak scale, in our analysis.  When the top
quark Yukawa coupling is large at the GUT scale,  the initial value of $A_t$ at
$M_{\ss GUT}$ tends to be  damped as $A_t$ is evolved, and  $m_{1/2}$ sets the
scale of $A_{t}$ at the weak scale.   $A_{t}$ is therefore set equal to its
quasi-fixed point value $ \approx  2m_{1/2} $ \cite{COPW}.  
  Here $m_{1/2}$ is the common gaugino mass at the
GUT scale and is equal to $4M_{2} \pi \alpha_{GUT} / g^{2}$, where we've
taken $1/\alpha_{\ss GUT}=24.85$.
Finally, the sfermion mass$^{2} $ matrix is diagonalized by a matrix $\eta$,
$\eta M^2 \eta^{-1}$, and can be parameterized by an angle $ \theta_{f} $ with

\[ \hspace{1cm} \left( \begin{array}{cc}
\cos{\theta_{f}} & \sin{\theta_{f}} e^{i \gamma_{f}} \nonumber \\
-\sin{\theta_{f}} e^{-i \gamma_{f}} & \cos{\theta_{f}}
\end{array} \right) 
\hspace{0.5cm}
 \equiv 
\hspace{0.5cm}
 \left( \begin{array}{cc}
\eta_{11} & \eta_{12} \nonumber \\
\eta_{21} & \eta_{22}
\end{array} \right)  \]

{}From the MSSM Lagrangian, we deduce the low energy effective four-fermi Lagrangian:
\begin{equation}
L = \bar{\chi} \gamma^\mu \gamma^5 \chi \bar{q_{i}} 
\gamma_{\mu} (\alpha_{1i} + \alpha_{2i} \gamma^{5}) q_{i} +
\alpha_{3i} \bar{\chi} \chi \bar{q_{i}} q_{i} + 
\alpha_{4i} \bar{\chi} \gamma^{5} \chi \bar{q_{i}} \gamma^{5} q_{i}+
\alpha_{5i} \bar{\chi} \chi \bar{q_{i}} \gamma^{5} q_{i} +
\alpha_{6i} \bar{\chi} \gamma^{5} \chi \bar{q_{i}} q_{i}
\label{lagr}
\end{equation}
The Lagrangian is to be summed over the quark generations, and the 
subscript i refers to up-type quarks (i=1) and down-type quarks
(i=2).  The coefficients $\alpha$ are: 

\begin{eqnarray}
\alpha_{1i} & = &  - \frac{1}{4(m^{2}_{1i} - m^{2}_{\chi})} \left[ 
\left| Y_{i} \right|^{2} - \left| X_{i}  \right|^{2} \right] 
- \frac{1}{4(m^{2}_{2i} - m^{2}_{\chi})} \left[
\left| V_{i} \right|^{2} - \left| W_{i} \right|^{2} \right]\nonumber \\
& & \mbox{} + \frac{g^{2}}{4m_{z}^{2}\cos^{2}{\theta_{w}}} \left[
\left| Z_{\chi_{3}} \right|^{2} - \left| Z_{\chi_{4}} \right|^{2}
\right] \frac{T_{3i} - 2 e_{i} \sin^{2}{\theta_{w}}}{2}
\label{alpha1}
\end{eqnarray}

\begin{eqnarray}
\alpha_{2i} & = & \frac{1}{4(m^{2}_{1i} - m^{2}_{\chi})} \left[
\left| Y_{i} \right|^{2} + \left| X_{i} \right|^{2} \right] 
+ \frac{1}{4(m^{2}_{2i} - m^{2}_{\chi})} \left[ 
\left| V_{i} \right|^{2} + \left| W_{i} \right|^{2} \right] \nonumber \\
& & \mbox{} - \frac{g^{2}}{4 m_{z}^{2} \cos^{2}{\theta_{w}}} \left[
\left| Z_{\chi_{3}} \right|^{2} - \left| Z_{\chi_{4}} \right|^{2}
\right] \frac{T_{3i}}{2}
\label{alpha2}
\end{eqnarray}

\begin{eqnarray}
\alpha_{3i} & = & - \frac{1}{2(m^{2}_{1i} - m^{2}_{\chi})} Re \left[
\left( X_{i} \right) \left( Y_{i} \right)^{\ast} \right] 
- \frac{1}{2(m^{2}_{2i} - m^{2}_{\chi})} Re \left[ 
\left( W_{i} \right) \left( V_{i} \right)^{\ast} \right] \nonumber \\
& & \mbox{} - \frac{g m_{qi}}{4 m_{w} B_{i}} \left[ Re \left( 
\zthree \right) D_{i} C_{i} \left( - \frac{1}{m^{2}_{H_{1}}} + 
\frac{1}{m^{2}_{H_{2}}} \right) \right. \nonumber \\
& & \mbox{} +  Re \left. \left( \zfour \right) \left( 
\frac{D_{i}^{2}}{m^{2}_{H_{2}}}+ \frac{C_{i}^{2}}{m^{2}_{H_{1}}} 
\right) \right]
\label{alpha3}
\end{eqnarray}

\begin{eqnarray}
\alpha_{4i} & = & - \frac{1}{2(m^{2}_{1i} - m^{2}_{\chi})} Re \left[
\left( X_{i} \right) \left( Y_{i} \right)^{\ast} \right]
- \frac{1}{2(m^{2}_{2i} - m^{2}_{\chi})} Re \left[ 
\left( W_{i} \right) \left( V_{i} \right)^{\ast} \right] \nonumber \\
& & \mbox{} - \frac{ g m_{qi} A_{i}}{4 m_{w} B_{i} m_{H_{3}}^{2}} \left[
\mbox{} - B_{i} Re \left( \zthree \right) \right. \nonumber \\
& & \left. \mbox{} + A_{i} Re \left( \zfour \right) \right] 
\label{alpha4}
\end{eqnarray}

\begin{eqnarray}
\alpha_{5i} & = & - \frac{i}{2(m^{2}_{1i} - m^{2}_{\chi})} Im \left[
\left( Y_{i} \right) \left( X_{i} \right)^{\ast} \right] 
- \frac{i}{2(m^{2}_{2i} - m^{2}_{\chi})} Im \left[
\left( V_{i} \right) \left( W_{i} \right)^{\ast} \right] \nonumber \\
& & \mbox{} + \frac{i g m_{qi} A_{i}}{4 m_{H_{3}}^{2} m_{w} B_{i}}
\left[ B_{i} Im \left( \zthree \right) \right. \nonumber \\
& & \mbox{} - \left. A_{i} Im \left( \zfour \right) \right]
\label{alpha5}
\end{eqnarray}

\begin{eqnarray}
\alpha_{6i} & = & - \frac{i}{2(m^{2}_{1i} - m^{2}_{\chi})} Im \left[
\left( Y_{i} \right) \left( X_{i} \right)^{\ast} \right] 
- \frac{i}{2(m^{2}_{2i} - m^{2}_{\chi})} Im \left[
\left( V_{i} \right) \left( W_{i} \right)^{\ast} \right] \nonumber \\
& & \mbox{} - \frac{i g m_{qi}}{4 m_{w} B_{i}} \left[ Im \left( 
\zthree \right) D_{i} C_{i} \left( - \frac{1}{m^{2}_{H_{1}}} + 
\frac{1}{m^{2}_{H_{2}}} \right) \right. \nonumber \\
& & \mbox{} +  Im \left. \left( \zfour \right) \left( 
\frac{D_{i}^{2}}{m^{2}_{H_{2}}}+ \frac{C_{i}^{2}}{m^{2}_{H_{1}}} 
\right) \right] 
\label{alpha6}
\end{eqnarray}

Where
\begin{eqnarray}
X_{i}& =& \Xi \nonumber \\
Y_{i}& =& \Yi \nonumber \\
W_{i}& =& \Wi \nonumber \\
V_{i}& =& \Vi
\label{xywz}
\end{eqnarray}
Here, $\delta_{1i}$ is $Z_{\chi 3}$ ($Z_{\chi 4}$), $\delta_{2i}$ is $Z_{\chi 4}$ ($-Z_{\chi 3}$),
$ B_{i} $ is $ \sin{\beta} $ ($ \cos{\beta} $), $ A_{i} 
$ is $ \cos{\beta} $ ($ -\sin{\beta} $), $ C_{i} $ is 
$ \sin{\alpha} $ ($ \cos{\alpha} $), and $ D_{i} $ is 
$ \cos{\alpha} $ ($ - \sin{\alpha} $) for up (down) type quarks.  The masses, $m_1$
and $m_2$ correspond to the two squark mass eigenstates. $m_{H_{1,2,3}}$ are the
two scalar and pseudoscalar Higgs masses respectively ($H_2$ is the lighter
scalar).
$ \alpha $ is the Higgs mixing angle.   We note that (\ref{alpha3})
corrects an error in \cite{FFO}, where the expressions for $\alpha_2$ and
$\alpha_3$ were presented previously. $y_i$ is the hypercharge defined by
$e_i = T_{3i} + y_i/2$.
The expressions for the $\alpha_i$ agree  
with those in \cite{CIN}.\footnote{Our expressions agree with the corrected
expressions in a revised version of \cite{CIN} as communicated to us by the
authors.}    Also, in the  limit of vanishing CP-violating phases, the
spin-independent term, proportional to
$\alpha_{3i}$, and the spin-dependent term, proportional to $\alpha_{2i}$, agree
with those in
\cite{review} and
\cite{ef}.

{}From the Lagrangian we can calculate the cross-section of a neutralino 
scattering off of a target nucleus.  All the terms in the Lagrangian,
however, do not contribute equally to the cross-section.  For example, 
\begin{equation}
\left| \langle f | \alpha_{4i} \bar{\chi} \gamma_{5} \chi \bar{q_{i}} 
 \gamma_{5} q_{i} |i \rangle \right|^{2} \propto |\zeta_{5i}|^2|
|\alpha_{4i}|^{2}  \frac{( - p_{\chi_{f}}\cdot p_{\chi_{i}} + m_{\chi}^{2} )}
{m_{\chi}^{2}} \frac{ ( - p_{N_{f}}\cdot p_{N_{i}} + m_{N}^{2} ) } {m_{N}^{2}}
\label{negligible3}
\end{equation}

\begin{equation}
\left| \langle f | \alpha_{5i} \bar{\chi}  \chi \bar{q_{i}} \gamma_{5} 
q_{i} |i \rangle \right|^{2} \propto |\zeta_{5i}|^2|
|\alpha_{5i}|^{2}  \frac{( - p_{N_{f}}\cdot p_{N_{i}} + m_{N}^{2} )}{m_{N}^{2}}
\label{negligible2}
\end{equation}

\begin{equation}
\left| \langle f | \alpha_{6i} \bar{\chi} \gamma^{5} \chi \bar{q_{i}} 
q_{i} |i \rangle \right|^{2} \propto
|\zeta_i|^2|\alpha_{6i}|^{2}  \frac{( - p_{\chi_{f}}\cdot p_{\chi_{i}} + m_{\chi}^{2} )}
{m_{\chi}^{2}}
\label{negligible1}
\end{equation}
where the $p_i$ are the four-momenta of the particle $i$.
In the non-relativistic limit, the expressions
(\ref{negligible3})-(\ref{negligible1})
have a leading term proportional to the
three momentum of the neutralino squared, the three momentum of the
nucleus squared, or both.  Here, $\zeta_{i} = \langle N | \bar{q_{i}} 
q_{i} | N \rangle $, and  $\zeta_{5i} = \langle N | \bar{q_{i}} 
 \gamma_{5} q_{i} | N \rangle $,  which corresponds to the quark contribution to
the nucleon spin and is discussed in detail in
\cite{formfactor}.  
The 
axial-vector term represented by $ \alpha_{1i} $ is also found to be
negligible
\cite{ef2}
\begin{eqnarray}
\lefteqn{ \left| \langle f | \alpha_{1i} \bar{\chi} \gamma_{\mu} \gamma_{5} 
\chi \bar{q_{i}}  
 \gamma_{\mu} q_{i} |i \rangle \right|^{2} \propto} \nonumber \\
& & |\alpha_{1i}|^{2}  \frac{ p_{\chi_{f}} p_{N_{f}} 
(p_{\chi_{i}} p_{N_{i}}) +
 p_{\chi_{f}} p_{N_{i}} (p_{\chi_{i}} p_{N_{f}})  
+ p_{N_{f}} p_{N_{i}}m_{\chi}^{2} - p_{\chi_{f}} p_{\chi_{i}}m_{N}^{2}
   - 2 m_{\chi}^{2} m_{N}^{2} }
{m_{\chi}^{2} m_{N}^{2}} 
\label{negligible4}
\end{eqnarray}
In the non-relativistic limit, this reduces to a combination of the 
three momentum squared of the neutralino and nucleus.
These terms are, hence, negligible compared to the terms multiplied by 
$\alpha_{2i}$ and $\alpha_{3i}$ which are not directly proportional to any
three-momenta in the non-relativistic limit
\cite{griest}.

The expressions for ${\alpha_2}_i$ and ${\alpha_3}_i$ in
equations (\ref{alpha2}) and 
(\ref{alpha3}) therefore, dominate
the elastic cross-section. 
These expressions receive contributions arising from squark,
$Z^{o}$, and both scalar Higgs boson exchanges.
The term proportional to ${\alpha_2}_i$ (equation
(\ref{alpha2})) is a spin dependent coupling of the LSP to  
the nucleus due to the $\gamma_{\mu} \gamma_{5}$ 
factor, which is  the spin projection in the non relativistic limit
\cite{ef2}.   On the other hand, the term proportional to
${\alpha_3}_i$ (equation (\ref{alpha3})) results in a coherent scattering of
the LSP with the nucleus and as such is proportional to the mass of the
nucleus.  For large $A$ nuclei, this term may be large for much of the
MSSM parameter space.   The steps to finding the cross-section from the
effective Lagrangian are laid out in
\cite{review} and outlined  below.

\section{Evaluating the Cross-Section}

The spin dependent cross-section can be written as
\begin{equation}
\sigma_{2} = \frac{32}{\pi} G_{f}^{2} m_{r}^{2} \Lambda^{2} J(J + 1)
\label{sd}
\end{equation}
where $m_{r}$ is the reduced neutralino-nucleus mass, $J$ is the spin 
of the nucleus and
\begin{equation}
\Lambda = \frac{1}{J} (a_{p} \langle S_{p} \rangle + a_{n} \langle
S_{n} \rangle)
\label{lamda}
\end{equation}
where
\begin{equation}
a_{p} = \sum_{i} \frac{\alpha_{2i}}{\sqrt{2} G_{f}} \Delta_{i}^{(p)}, 
a_{n} = \sum_{i} \frac{\alpha_{2i}}{\sqrt{2} G_{f}} \Delta_{i}^{(n)}
\label{a}
\end{equation}
The factors $\Delta_{i}^{(p,n)}$ depend on the spin content of the nucleon
and are taken here to be $\Delta_{i}^{(p)} = 0.77, -0.38, -0.09 $ for 
$u, d, s$ respectively and $\Delta_{u}^{(n)} = \Delta_{d}^{(p)}, 
\Delta_{d}^{(n)} = \Delta_{u}^{(p)}, \Delta_{s}^{(n)} =\Delta_{s}^{(p)}$
\cite{adams}  The expectation values of the spin content of the nucleus, 
$ \langle S_{p,n} \rangle $, depend on the target nucleus.  Our results 
are for $^{73}$Ge, which has $ \langle S_{p,n} \rangle = 0.011, 0.491 $,
and for $^{19}$F, which has $ \langle S_{p,n} \rangle = 0.415, -0.047 $.
The reader may refer to \cite{review} for details on these quantities. 

The spin-independent cross-section can be written as
\begin{equation}
\sigma_{3} = \frac{4 m_{r}^{2}}{\pi} \left[ Z f_{p} + (A-Z) f_{n} 
\right]^{2}
\label{si}
\end{equation}
where
\begin{equation}
\frac{f_{p}}{m_{p}} = \sum_{q=u,d,s} f_{Tq}^{(p)} 
\frac{\alpha_{3q}}{m_{q}} +
\frac{2}{27} f_{TG}^{(p)} \sum_{c,b,t} \frac{\alpha_{3q}}{m_q}
\label{f}
\end{equation}
and $f_{n}$ has a similar expression.  The parameters $f_{Tq}^{(p)}$ 
are defined
by $ \langle p | m_{q} \bar{q} q | p \rangle = m_{p} f_{Tq}^{(p)} $, 
while $ f_{TG}^{(p)} = 1 - \sum_{q=u,d,s} f_{Tq}^{(p)} $ \cite{SVZ}.  Our 
results used $ f_{Tq}^{(p)} = 0.019, 0.041, 0.14 $ for $ u,d,s $ and 
$ f_{Tq}^{(n)} = 0.023, 0.034, 0.14 $ \cite{GLS}.  There are other
contributions to the spin-independent cross-section that depend on loop
effects for heavy quarks and twist-2 operators which have been shown to
be numerically small \cite{DN} and are not included here.  Note from
(\ref{alpha3}) and (\ref{xywz}) that
the spin-independent cross-section is suppressed by the small
nucleon mass.
However, in contrast to the spin-dependent cross-section, the
spin-independent cross-section scales with the atomic weight of
the scattering nucleus and therefore can dominate for heavy nuclei. 
We show results for $^{73}$Ge, which has
predominantly spin-independent interactions, and 
$^{19}$F, for which either interaction can dominate, depending on the
parameter region.

The CP-violating phases can impact the cross-section in several ways.
The phases can change the mass of the LSP.  They will also change 
$Z_{\chi i}$, which indicate the identity of the LSP, as well as the
$\eta_{ij}$, which indicate the mass eigenstates (and eigenvalues) of the
sfermions.  To get an understanding of how these effects arise, consider
the case of  the LSP being a nearly pure bino.  (The neutralino
can account for most of dark matter when it is mostly a bino.)  In this case, 
$Z_{\chi 1}$ is nearly 1 and the other $Z_{\chi i}$
 are very small.  Then, the spin dependent term is approximately

\begin{eqnarray}
\alpha_{2i} & \simeq & \frac{1}{4(m^{2}_{1i} - m^{2}_{\chi})} \left[
\left| \eta^{\ast}_{11}  \frac{y_{i}}{2} g' Z_{\chi 1} \right|^{2} 
+ \left| \eta_{12}^{\ast} e_{i} g' Z_{\chi 1}^{\ast} \right|^{2} \right] 
\nonumber \\
& & \mbox{} + \frac{1}{4(m^{2}_{2i} - m^{2}_{\chi})} \left[
\left| \eta^{\ast}_{21}  \frac{y_{i}}{2} g' Z_{\chi 1} \right|^{2} 
+ \left| \eta_{22}^{\ast} e_{i} g' Z_{\chi 1}^{\ast} \right|^{2} \right]
\end{eqnarray}
Hence, in this limit, the spin dependent cross-section will depend
on the phases primarily through their effect on the mass of the
neutralino, although the magnitude of the $\eta_{\,ij}$ will vary significantly, as
well, if $R_i|\mu|$ is comparable to $|A_i|$ (see (\ref{offdiag})).

To approximate the spin-independent term is somewhat more complicated due to
the light scalar Higgs exchange. Keeping the leading terms we find

\begin{eqnarray}
\alpha_{3i} & \simeq & - \frac{1}{2(m^{2}_{1i} - m^{2}_{\chi})} Re \left[
\left(  - \eta_{12}^{\ast} e_{i} g' Z_{\chi 1}^{\ast} \right) 
\left( \eta^{\ast}_{11} \frac{y_{i}}{2} g' Z_{\chi 1} \right)^{\ast}
\right] 
\nonumber \\
& & \mbox{} - \frac{1}{2(m^{2}_{2i} - m^{2}_{\chi})} Re \left[  
\left( - \eta_{22}^{\ast} e_{i} g' Z_{\chi 1}^{\ast} \right) 
\left( \eta_{21}^{\ast} \frac{y_{i}}{2} g' Z_{\chi 1} \right)^{\ast} 
\right]  \nonumber \\
& & \mbox{} + \frac{g g' m_{qi}}{4 m_{w} {m^{2}_{H_{2}}}  B_{i}} \left[ Re  
\left( 
\delta_{1i} Z_{\chi1} \right)  D_{i} C_{i} 
 +  Re  \left( \delta_{2i} Z_{\chi1} \right) 
{D_{i}^{2}}
\label{sie}
 \right]
\end{eqnarray}
Due to the relative size of the Higgs mass to the squark masses, and the smallness
of $\eta_{12}$ to $\eta_{11}$, the Higgs exchange term can dominate in
$\alpha_3$.
  In this limit, the spin-independent cross-section will depend on
the phases through the mass of the neutralino, the
phases in $Z_{\chi 1}$, and the phases in $Z_{\chi 3,4}$ which exhibit a strong
dependence on $\theta_\mu$.   It would be expected, then, that 
the spin-independent term should be more sensitive to the changes in 
the CP phases as we will show below.  We have not included the effect of the CP violating 
phases on the light Higgs mass \cite{pw}, which will typically be shifted by
a few GeV for the masses and phases we study.

\section{Results}

Using the analytic expressions for the cross-section, we first show a set
of contour plots on the $M_{2} - \mu$ plane for specific values of 
$ A_{f} $, $M_{L(R)}$, $ \tan{\beta} $, and $ \theta_{A} $. In Figures 1 --
6, we display the total cross-section, spin-dependent cross-section, and
spin-independent cross-section for both $^{73}$Ge (Figs. 1 -- 3) and
$^{19}$F (Figs. 4 -- 6).  In these plots, the electric dipole moment
constraints  of the electron, neutron and mercury atom have not been
imposed in order to show the  general behavior of the cross-section as
$\theta_{\mu}$ changes.   The second set
of plots (Figs. 7 -- 10) are scatter plots over the MSSM parameter space
projected onto the 
$M_{2} - \mu$ plane, concentrating on regions where the electric
dipole constraints are satisfied.  
These plots serve as an existence proof
that new CP violating phases can have a significant effect on the
scattering rates while still satisfying the EDM constraints.
Again, separate scatter plots will be made for $^{73}$Ge and $^{19}$F.

\subsection{General Behavior}
\label{sec-general}

In Figs. 1 -- 6, we set $A_f=3000$ GeV, $\tan{\beta} = 3$, 
$m_{0}=100$ GeV, $m_A=300$ GeV and $ \theta_{A}=\pi/2$.  
$ A_{t}$ is set to its quasi-fixed point value $\simeq 2
m_{1/2}$.  The
value of $A_f$ is chosen for easy comparison with Figures 7--8, where
large $A_f$ is necessary to satisfy the electric dipole moments constraints. 
 Figures 1a--c
show how the total cross-section  for scattering off of $^{73}$Ge
changes as $\theta_{\mu}$ 
changes from 0 to $\pi/4$.  The most significant change 
occurs in the lower right quadrant.  To determine what causes 
this change, the spin-independent and spin-dependent cross-sections
are plotted separately in Figs. 2 and 3.  Direct detection
experiments typically are sensitive to primarily one or the other type of
interaction, and it is important to understand how the phases affect
the different cross-sections.
In Figs.~2a--c, we see that 
the spin-dependent cross-section varies only very mildly 
with $\thm$.   In contrast, Figs.~3a--c
show that the spin-independent cross-section (which has a maximum near
the line $\mu=M_2/2$, where the bino-Higgsino mixing is large) for
$^{73}$Ge does change significantly, by up to a factor of 2, as $\theta_{\mu}$ 
is varied from 0 to $\pi/4$, and the effect is largest 
in the lower right quadrant of the plane.  The total cross-section's
change is due to the large change in the spin-independent cross-section.

Figures 4--6 are the corresponding pictures for $^{19}$F.  
Again, the spin-dependent cross-section does not change much as 
$\theta_{\mu}$ increases, while the spin-independent term undergoes a 
significant change.  Since $^{19}$F is so 
light, the spin-independent term is comparable to the spin-dependent
term over much of the parameter space. The total cross-section 
is affected by the change in $\thm$ in those regions where the
spin-independent scattering cross-section dominates (for large $|\mu|$
and small $M_2$) and only little affected in the regions where the
spin-dependent scattering cross-section dominates (for large $M_2$
and small $|\mu|$).

\subsection{Behavior with EDM Constraints}
 
When the CP-violating phases are non-zero, the MSSM
predicts a non-zero electric dipole moment for the electron, neutron and the
mercury atom.  In fact, over much of the parameter space, the 
values chosen in the previous plots will not satisfy the experimental 
bounds for the electric dipole moments 
$( < 10^{-25} e$ cm \cite{nedm} for the neutron) ,  $( < 4.2 \times 10^{-27} e$ cm
\cite{eedm} for the electron), and $( < 9 \times 10^{-28} e$ cm \cite{medm} for the
mercury atom). 

Within the MSSM, there are several contributions to the electric 
dipole moment.  For the leptons and quarks, electric dipole moments are 
induced at one loop via exchange of sfermions, charginos, neutralinos, or
gluinos (for quarks).  For the neutron, there are additional contributions to the 
electric dipole moment, beyond the contribution by the quark EDMs, which 
include a gluonic contribution and a quark color dipole contribution
\cite{KO,wads,IN2}.  For a given set of values of the CP-violating phases, 
there may be significant cancellations between the individual
contributions \cite{fko1,IN2,fko2}.
In general, 
the electric dipole moment of the electron provides a more stringent
constraint over the $M_{2} - \mu$ plane than does the neutron electric
dipole moment, especially when uncertainties in the computation of $d_n$ are
taken into account \cite{fopr}.
It is however, important to have more than one independent experimental constraint on
edms, in order to obtain limits on the two CP violating phases.  In \cite{fopr}, it was
argued that the edm of $^{199}$Hg (induced through T-violating nuclear forces
\cite{krip}) could be used to constrain the MSSM phases.
  Although it may require some degree of
fine tuning, for any given set of $\tan{\beta},
\mu, M_{2}, m_{0}, \theta_{A_i},$ and $\theta_{\mu}$
it is always possible to satisfy both of 
the electric dipole moments by choosing     
large enough magnitudes for the trilinear mass parameters, $A_i$
\cite{garisto,fko2,aad} (more correctly, acceptable solutions 
exist for either $\theta_A$ or $\theta_{A}\rightarrow\theta_{A}+\pi$).  
In the absence of tuning the sfermion masses, the
phases must take on  values of order $10^{-2}-10^{-3}$ in order to satisfy the EDM
constraints.  

For figures 7 -- 10, a general scan of parameter space was performed 
over the $M_{2}, \mu, A,$ and $m_{0}$ while setting $A_{t}$ again equal
to the quasi-fixed point. We have again taken $\tan \beta = 3$ and $m_A =
300$ GeV. The range of parameters is chosen via several criteria.  From
LEP2's search for charginos, there is a lower bound on  the mass of the
chargino of 95 GeV
\cite{lep}.  This implies a restriction on the lower values of $M_{2}$
and $\mu$ of around 100 GeV. We also insist that the  neutralino be the
LSP, and  we employ the recent LEP2 constraint
on sfermion masses and discard runs with $m_{\tilde t} < 80$ GeV \cite{lep}.  
Furthermore, when the
sfermions are very massive and the LSP is gaugino-like,
the neutralino's relic density is too large to be compatible with the
observational lower limits on the lifetime of the universe \cite{efos}. 
Therefore for simplicity, we restrict the values of 
$m_{0}$ so as to give cosmologically reasonable relic
densities for all neutralino compositions (although of course $m_0$ is 
not constrained for Higgsino-like LSP's).

In each figure, $\theta_{A}$ and $\theta_{\mu}$
are fixed, and a scan is performed over $M_2,|\mu|,A$, and $m_0$.  If a set of parameters satisfies the
EDM constraints, the value of the cross-section is calculated and 
compared to the value of the cross-section when  
$\theta_{A}$ and $\theta_{\mu}$ are both zero.  This scan is then
projected onto the $M_{2} - \mu$ plane.  Figs.~7a and 7b are for 
$^{73}$Ge for $\thm=\pi/8,\tha=3\pi/8$ and  $\thm=\pi/4,\tha=\pi/2$, respectively, 
and 8a and 8b are the corresponding figures for $^{19}$F. In Figures 7 and 8, we
require both the electron and neutron edm limits to be satisfied. For Figs.~7a and 8a,
the range of parameters is:  200 GeV
$\leq M_{2}
\leq $ 800 GeV, 100 GeV $\leq \mu \leq $ 1000 GeV, 100 GeV $\leq m_{0}
\leq $ 200 GeV, and 2200 GeV $\leq A \leq $ 3500 GeV.  In the
scan, $M_{2}$ is
incremented every 20 GeV, $\mu$ every 20 GeV, $m_{0}$ every 25 GeV, and
$A$ every 25 GeV.  
For figures 7b and 8b, the range of parameters is:  150 GeV $\leq M_{2}
\leq $ 400 GeV, 100 GeV $\leq \mu \leq $ 1000 GeV, 100 GeV $\leq m_{0}
\leq $ 200 GeV, and 2700 GeV $\leq A \leq $ 3500 GeV.  In the
scan, $M_{2}$ was 
incremented every 10 GeV, $\mu$ every 20 GeV, $m_{0}$ every 10 GeV, and
$A$ every 25 GeV.   We bin the resulting
$\sigma(\thm,\tha\neq 0)/\sigma(\thm,\tha=0)$ in steps of 0.1.  A
single symbol at a point means that every parameter set satisfying the
EDM constraints has a cross-section ratio in the specified range.
Some grid points have two symbols.
Notice for all of the diagrams, there are regions where the cross-section
is reduced.  For some of the diagrams, however, there are regions where the 
cross-section is enhanced.
In figures 7b and 8b, when $\theta_{\mu} = \pi/4$ and 
$\theta_{A} = \pi/2$ there are regions where the cross-section decreases
by up to a factor of 2.  Although this region of parameter space is small
(in the region of  200 GeV$ \leq M_{2} \leq $ 300 GeV
 and $\mu$ around 750 GeV), it
has cosmological significance because it 
is in a region of parameter space
where the neutralino is most nearly a bino, so that the neutralino
can account for most of dark matter.
By comparing
the scatterplots to the previous contour plots,
the largest changes identified by the scatterplots correspond
to the region of parameter space that has the smallest cross-section
for both $^{73}$Ge and $^{19}$F.

In \cite{FFO}, we had found cases in which the cross-section was reduced
by over an order of magnitude for certain values of the phases. 
Those results show the maximum possible effect and such a reduction
occurs at a different value of $\theta_\mu$ for each choice of 
$M_2$ and $\mu$.  Here, we have a milder effect, but one that should be
taken as relatively generic.  We have chosen two specific values of
$\theta_\mu$, and scan the $M_2$-$\mu$ plane.  Therefore, in most cases,
we will miss the value of $\theta_\mu$ for which the cancellation in the
cross-section is at a maximum.  Furthermore, we have lost a considerable
amount of parameter space due to the imposition of the EDM constraints.
This space could be enlarged if we were to relax the universality on the
$A$-terms. By adjusting $A_e$ and $A_q$ separately, a larger portion of
the plane would survive and larger effects would be visible.

 In Figures 9 and 10, we show the results when both the
electron and mercury edms are satisfied.
Figures 9a and 10a use the same range of parameters as those in figures
7a and 8a.  Likewise, figures 9b and 10b use the same parameters as those in 
figures 7b and 8b.  The variation of the cross-section in these plots is
very similar to the variation of the cross-section when the neutron 
and electron edms are imposed (figures 7 and 8) but a different section
of the $M_{2} - |\mu|$ plane is selected.  These results (in figures 7
-- 10) are in qualitative agreement with those of \cite{CIN}.

It is important to note that we have
chosen the ranges for our scans of $A$ to coincide with the regions
where cancellations are known to make the electric dipole moments
small.  As such, each point in Figs.7 -- 10a represents 
about 60 points in the scan (deprojected), and thus we find solutions
for about 12\% of the total of
$\sim$400,000 points in our scan. Similarly, each point in Figs.7 -- 10b
represents only about 10 points in the scan. In this case, we find
solutions for about 1\% of a total of
$\sim$400,000 points in the scan.  These points serve to demonstrate the
size of the uncertainties in the elastic scattering rates due to large CP
violating phases. Again, the cosmological constraints may be satisfied by
taking the first two sfermion generations heavy, while having a light
third generation\cite{heavy}.  Lastly, we also note that there are
regions at large $m_0 (>1{\rm TeV})$ where the phases can be large
without any tuning of the mass parameters.  However, such large sfermion
masses are cosmologically forbidden unless the LSP is mixed or
Higgsino-like, in which case there tends to be an insufficient relic LSP
abundance to make direct detection relevant.

\section{Summary}

We have made a comprehensive study of the effect of the two CP violating phases in the
MSSM on the neutralino elastic scattering cross-section. 
Starting with the full 4-Fermi interaction lagrangian between neutralinos and quarks,
we examined in detail the effect of the phases of the two dominant terms
corresponding to the spin-dependent ($\alpha_2$) and spin-independent ($\alpha_3$)
parts of the cross-section in the non-relativistic limit.

We have shown explicitly the variation of the cross-section in the $M_2$-$\mu$
plane for neutralino scattering off two sample nuclei, $^{73}$Ge and $^{19}$F.
We showed the extent of the variation for the total cross-section as well as the 
individual spin-dependent and -independent contributions.  It is the spin-independent
piece in the bino-portion of the parameter plane that is most sensitive to the 
CP-violating phases. 

We have also considered in detail the regions of the parameter plane which are
consistent with the experimental limits on the electric dipole moments of the
electron, neutron,  and mercury atom.   The edms provide a very strong constraint 
and therefore, some degree of tuning is necessary to satisfy these constraints 
along with other cosmological and phenomenological constraints as well. 
In the surviving regions, the cross-section is typically reduced, by as much as a factor
of $\sim$ 2.5, while some regions in the $M_2$-$\mu$ plane show slight enhancements in
the cross-section by about 10\%.

It is also worth mentioning that the relic density of neutralinos
is determined by their annihilation rate and the annihilation rate will
also depend on the CP violating phases.  In the general MSSM, these effect
could be large \cite{FOS} as the typical p-wave suppression of the annihilation
cross-section is removed when the left and right handed sfermion masses are nearly
degenerate.  In the models such as the constrained MSSM (when universal soft masses at
the GUT scale are assumed) this effect is small \cite{fko1}. 
There is however, another contribution to the annihilation cross-section
which is unique to the presence of non-vanishing CP-violating phases.
In the
effective Lagrangian (\ref{lagr}), terms which go as  
$\bar{\chi} \gamma^{5} \chi $ will be proportional to the three momentum
of the neutralinos and will thus be negligible in the 
non-relativistic limit.  However, the term 
proportional to $ \alpha_{5i} $ will give non-negligible contributions
to the annihilation rate for non-zero CP violating phases as can be seen by examination
of eq.(\ref{alpha5}).  In the  non-relativistic limit we have
\begin{equation}
\sigma v_{rel} = \sum_{f} \frac{2}{\pi} \left| \alpha_{5i} \right|^{2} 
m_{\chi}^{2} \left( 1 - \frac{m_{f}^{2}}{m_{\chi}^{2}} \right)^{\frac{3}{2}}
\label{annhilation}
\end{equation}
where the total contribution should be summed over all fermions, f, which
are less massive then the neutralino.  
Note that in the limit of 
vanishing CP violating phases, $\alpha_{5i}$  is zero since it
depends on the imaginary contribution to (\ref{xywz}).  
 A complete investigation of the effects of the phases on the total 
annihilation cross-section is left for future study.

\vskip .3in
\vbox{
\noindent{ {\bf Acknowledgments} } \\
\noindent  The work of K.O. was supported in part by DOE grant
DE--FG02--94ER--40823.  The work of T.F. was supported in part by DOE   
grant DE--FG02--95ER--40896 and in part by the University of Wisconsin  
Research Committee with funds granted by the Wisconsin Alumni Research  
Foundation.}

\newpage

\begin{figure}
\vspace*{-1.5in} 
\hspace*{-0.5in}
\begin{minipage}{7.5in}
\epsfig{file=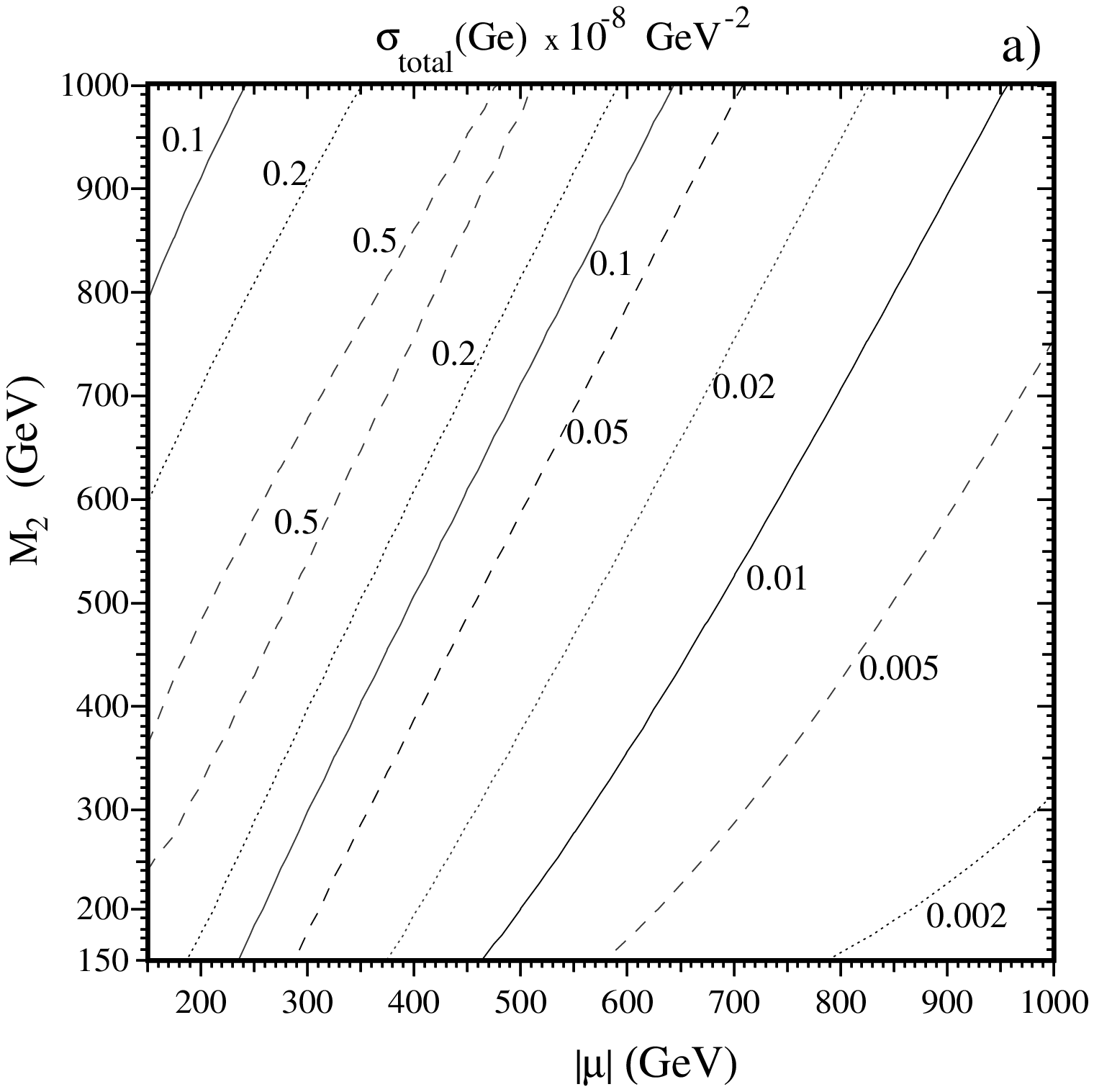,height=5.5in} 
 \hspace*{-0.8in}
\epsfig{file=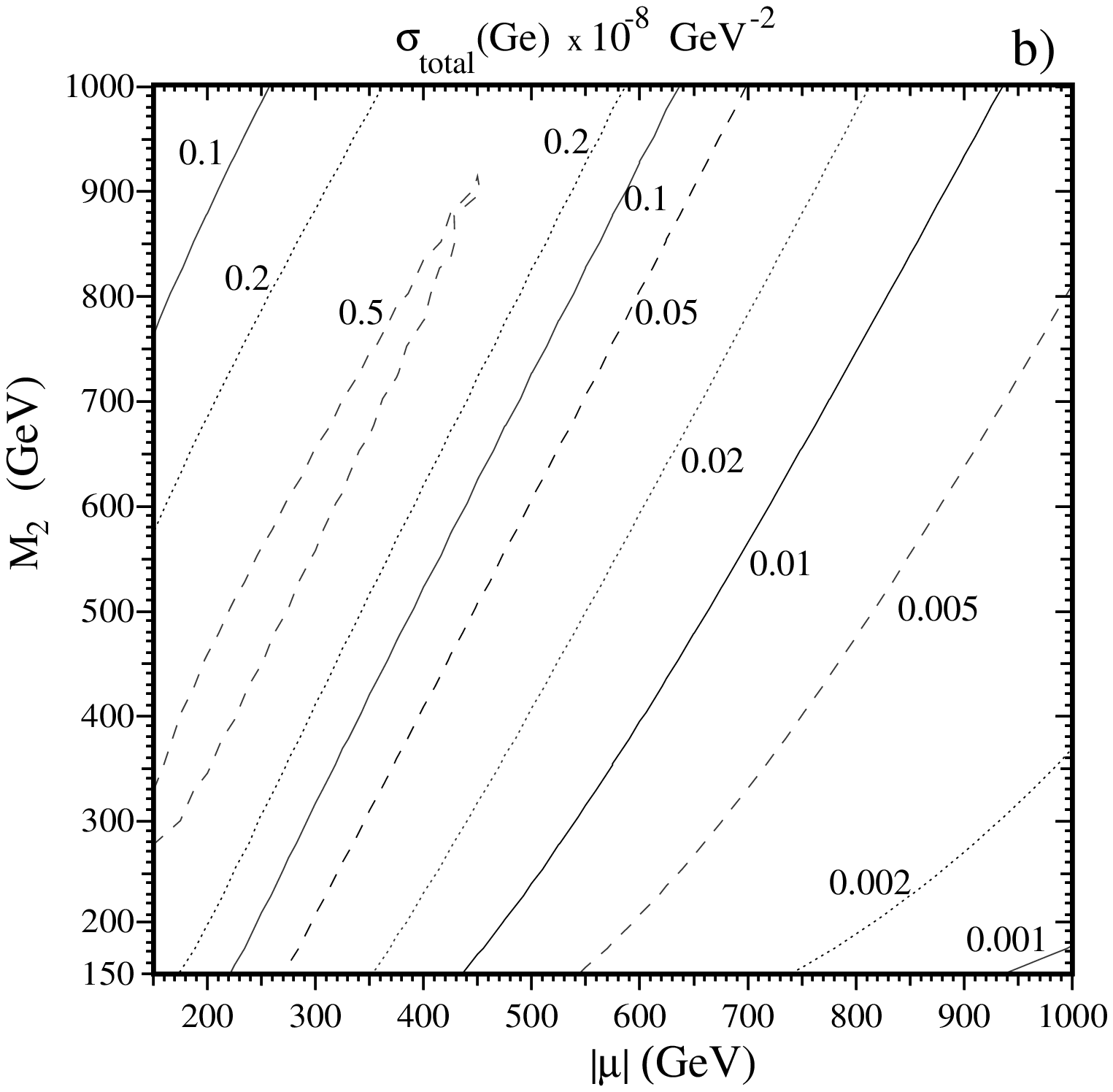,height=5.5in} 
\end{minipage}
\begin{minipage}{6.5in}
\vspace*{-1.9in}
\hspace{1.0in}
\epsfig{file=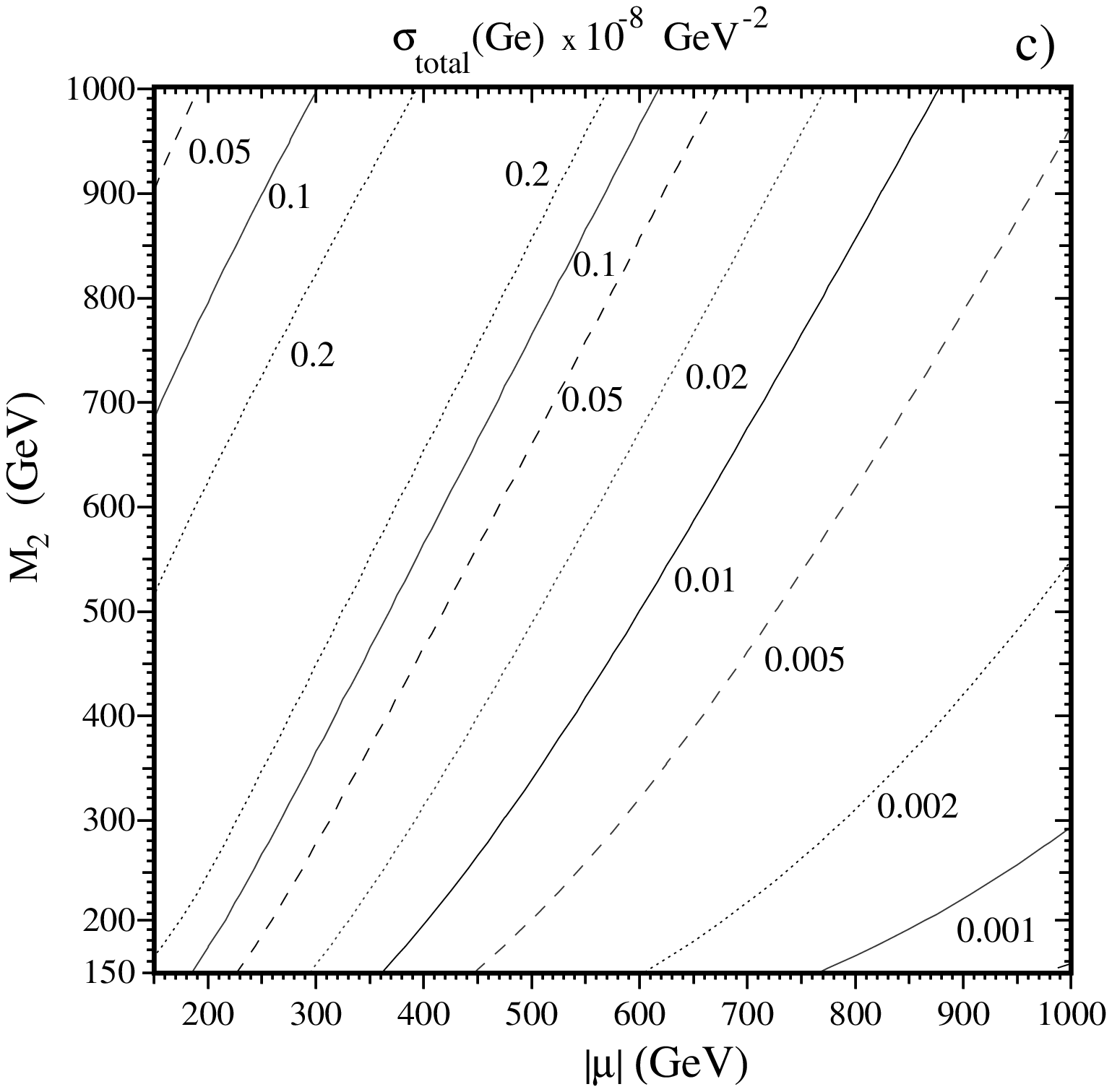,height=5.5in}
\end{minipage}
\vspace*{-0.6in}
\caption{\it The total cross-section for elastic scattering off of
$^{73}$Ge a.) for
$\theta_\mu = 0$, b) $\theta_\mu = \pi/8$, c.) $\theta_\mu = \pi/4$.  All plots have
$\tha = \pi/2$.  The axis are in units of GeV.  The contours are in units of $10^{-8}
$ GeV$^{-2}$. }
\end{figure}

\begin{figure}
\vspace*{-1.5in} 
 \hspace*{-0.5in}
\begin{minipage}{7.5in}
\epsfig{file=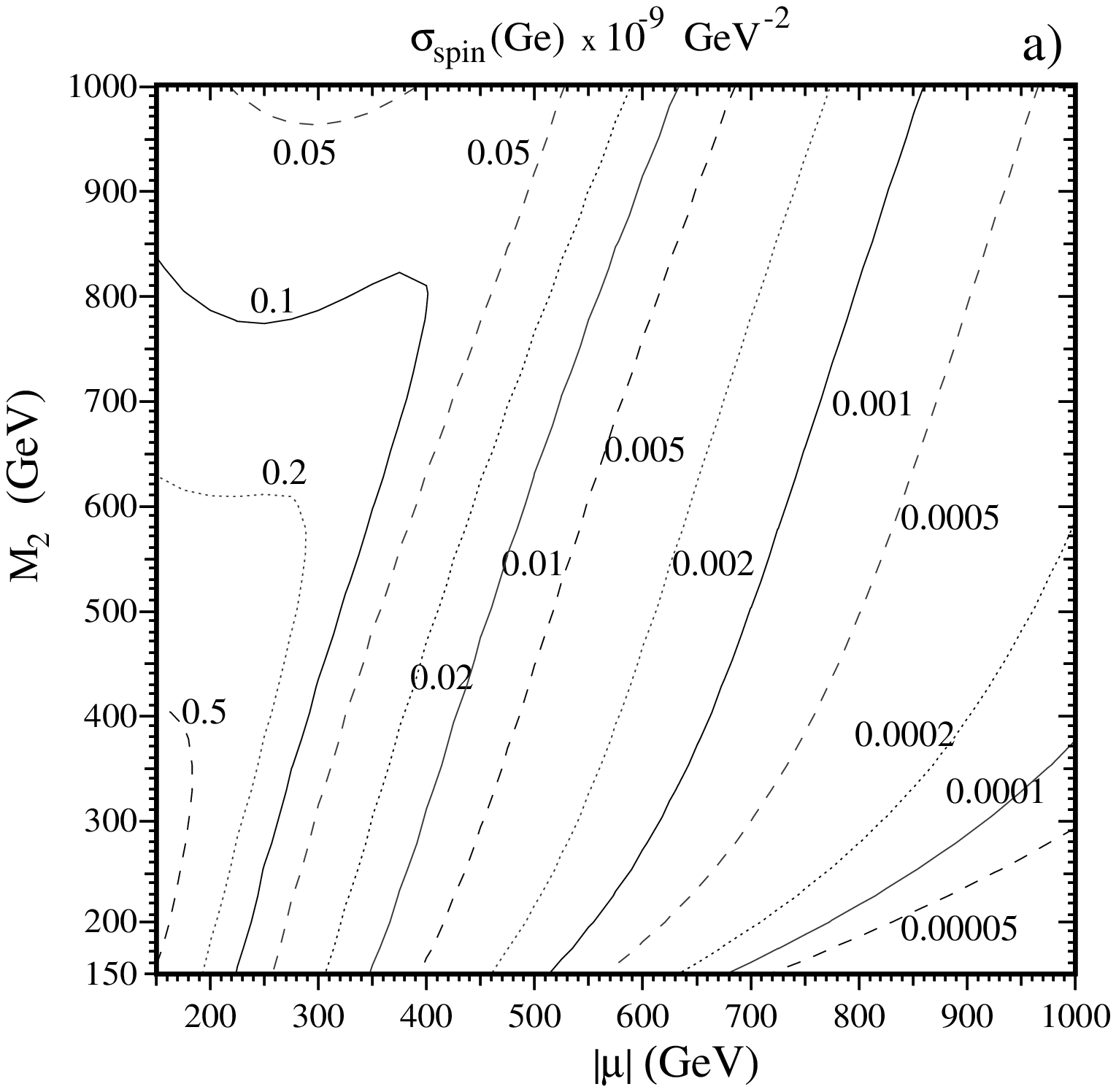,height=5.5in} 
 \hspace*{-0.8in}
\epsfig{file=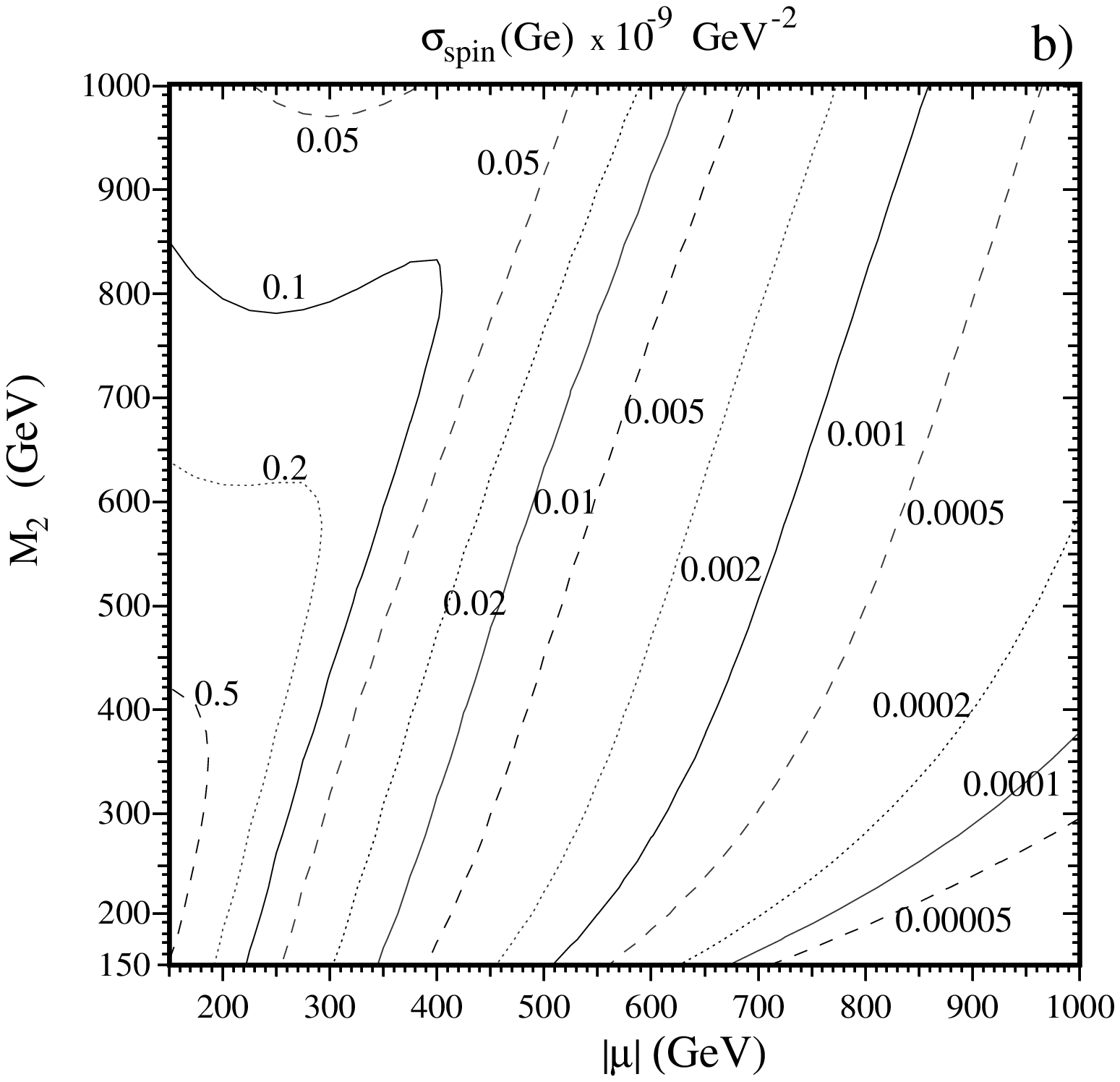,height=5.5in} 
\end{minipage}
\begin{minipage}{6.5in}
\vspace*{-1.9in}
\hspace{1.in}
\epsfig{file=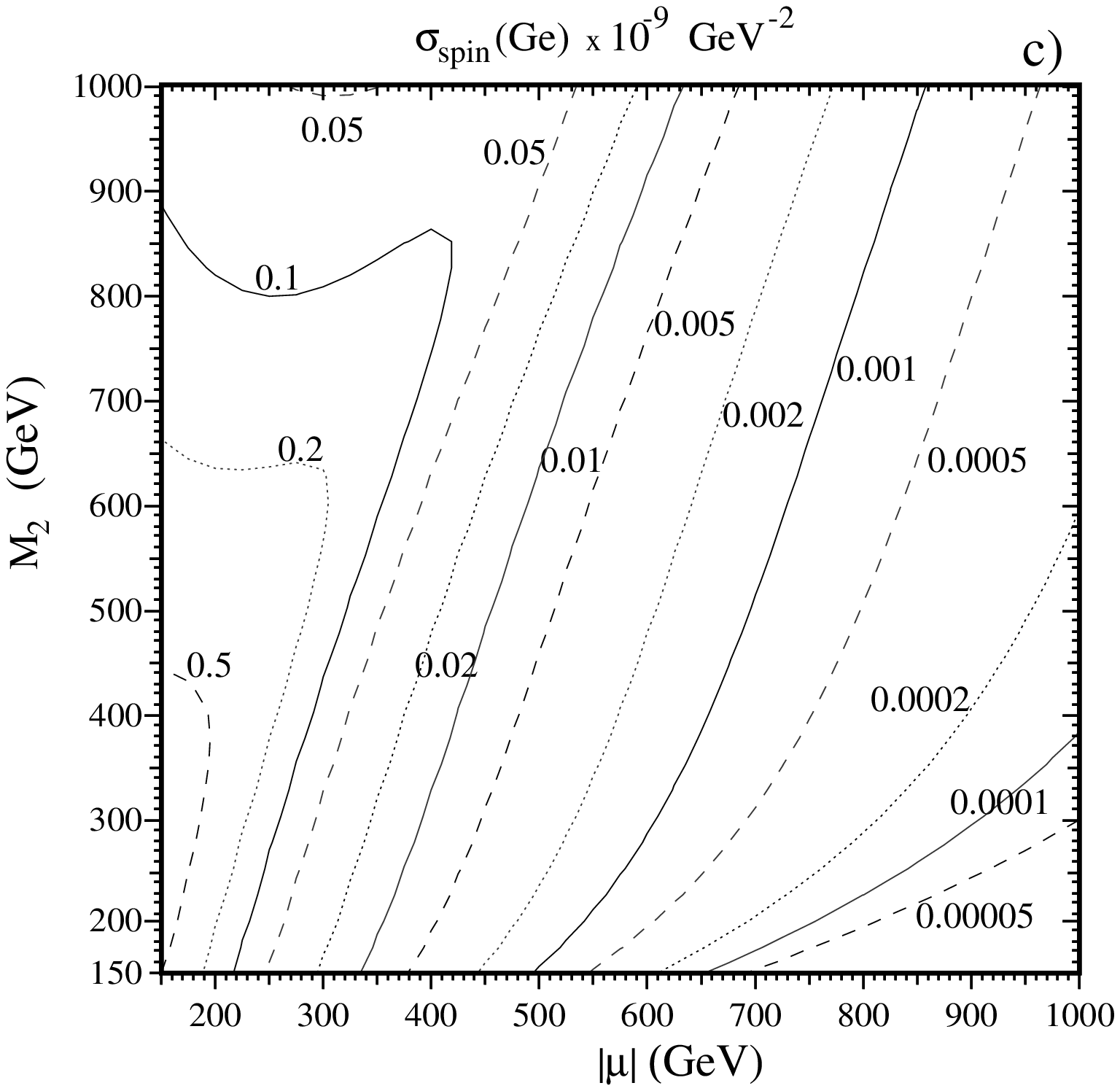,height=5.5in}
\end{minipage}
\vspace*{-0.6in}
\caption{\it The spin-dependent cross-section for elastic scattering
  off of $^{73}$Ge, for the same parameters as in Fig.~1. The contours
  are in units of $10^{-9}$ GeV$^{-2}$.}
\end{figure}

\begin{figure}
\vspace*{-1.5in} 
\hspace*{-0.5in}
\begin{minipage}{7.5in}
\epsfig{file=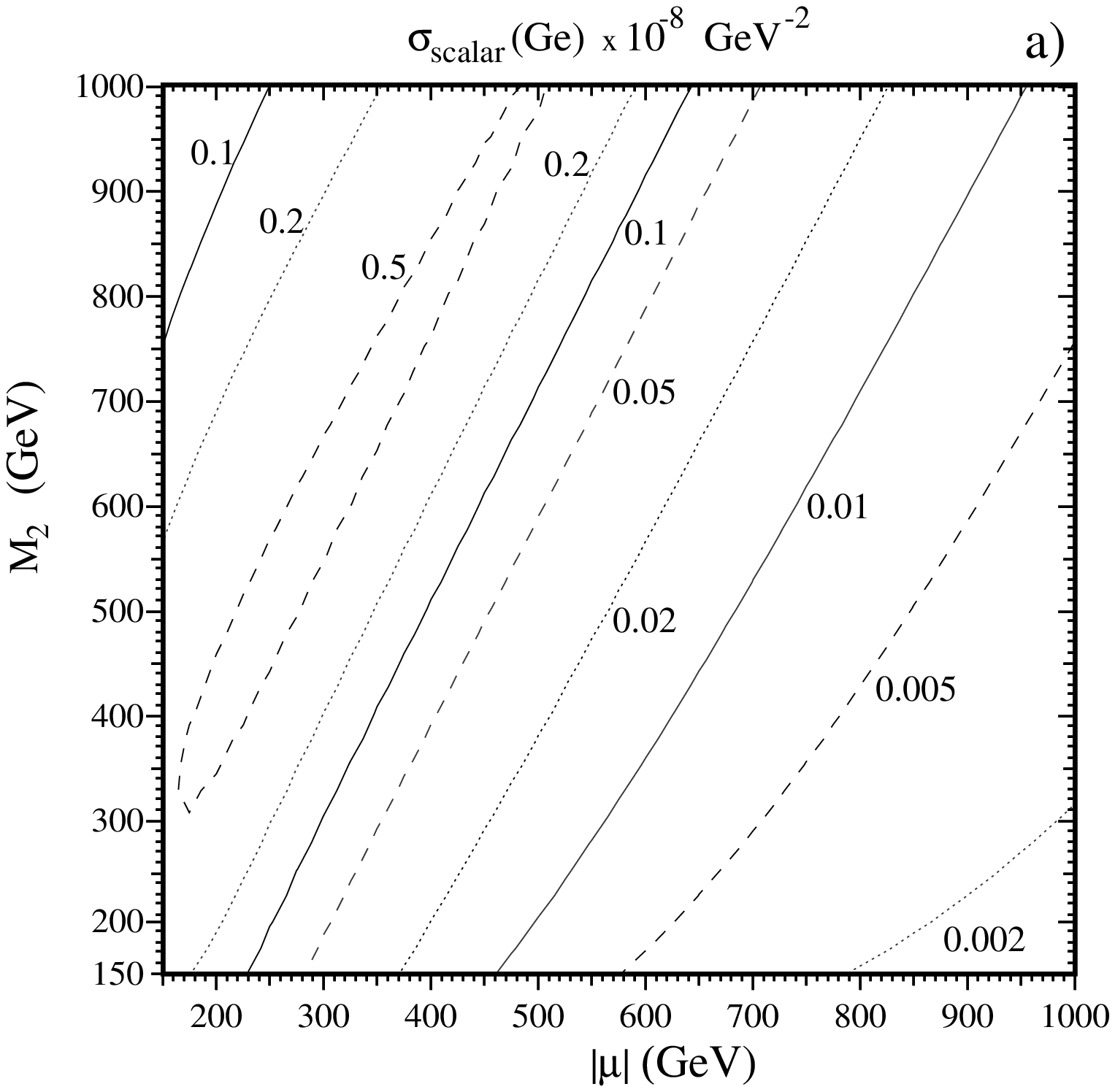,height=5.5in} 
 \hspace*{-0.8in}
\epsfig{file=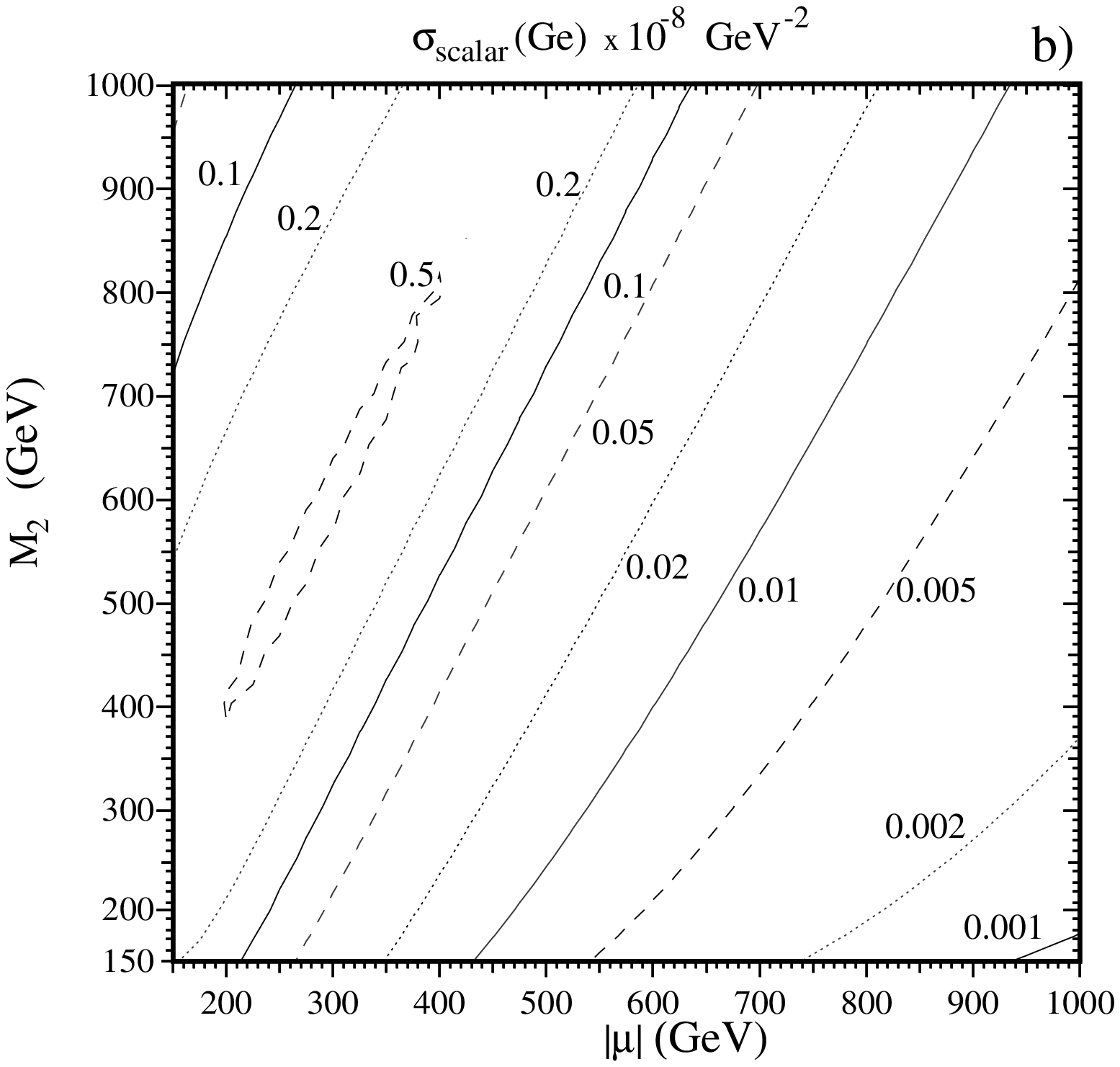,height=5.5in} 
\end{minipage}
\begin{minipage}{6.5in}
\vspace*{-1.9in}
\hspace{1.0in}
\epsfig{file=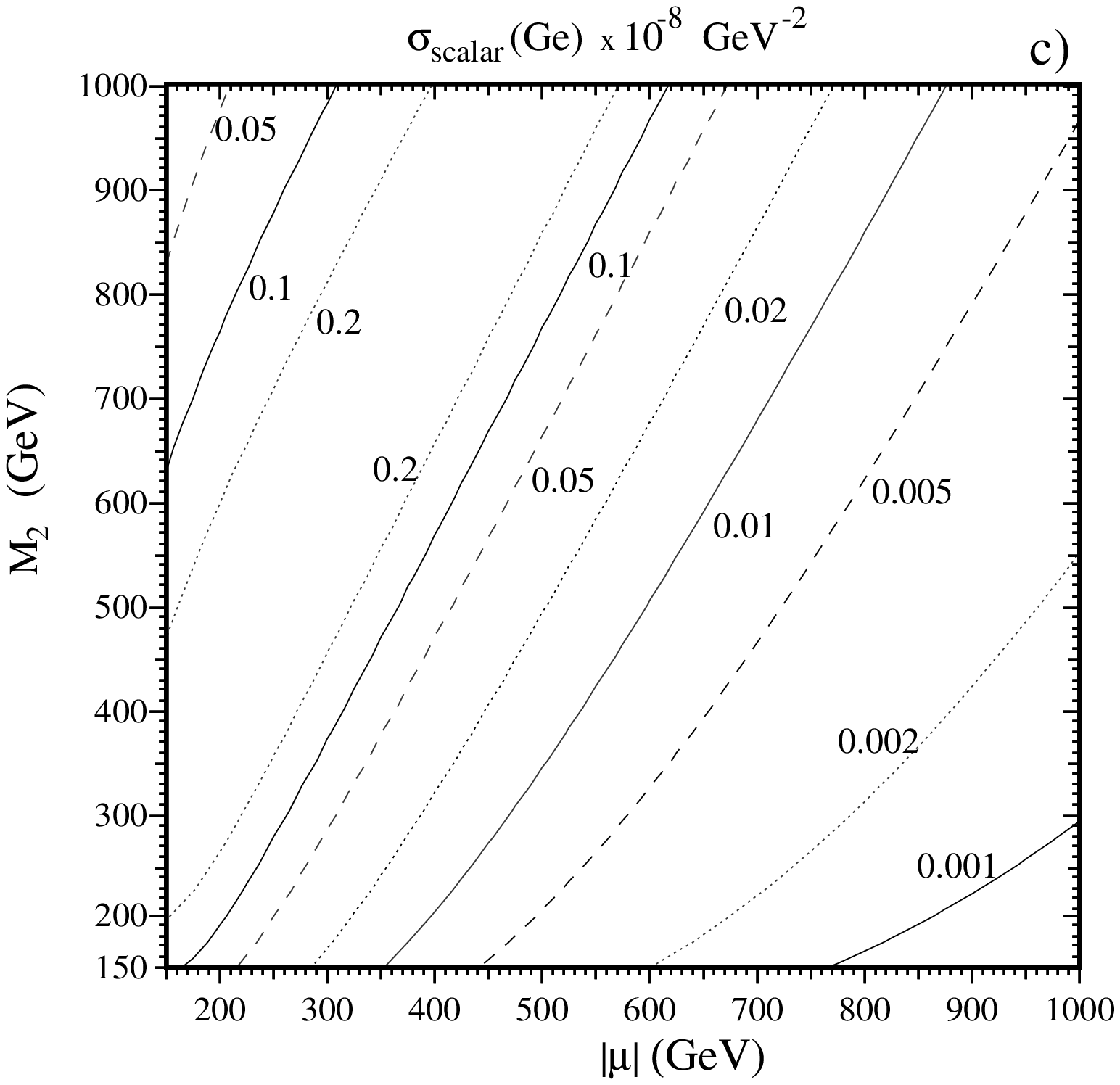,height=5.5in}
\end{minipage}
\vspace*{-0.6in}
\caption{\it The spin-independent cross-section for elastic scattering
  off of $^{73}$Ge, for the same parameters as in Fig.~1.  The contours
  are in units of $10^{-8}$ GeV$^{-2}$ }
\end{figure}

\begin{figure}
\vspace*{-1.5in} 
\hspace*{-0.5in}
\begin{minipage}{7.5in}
\epsfig{file=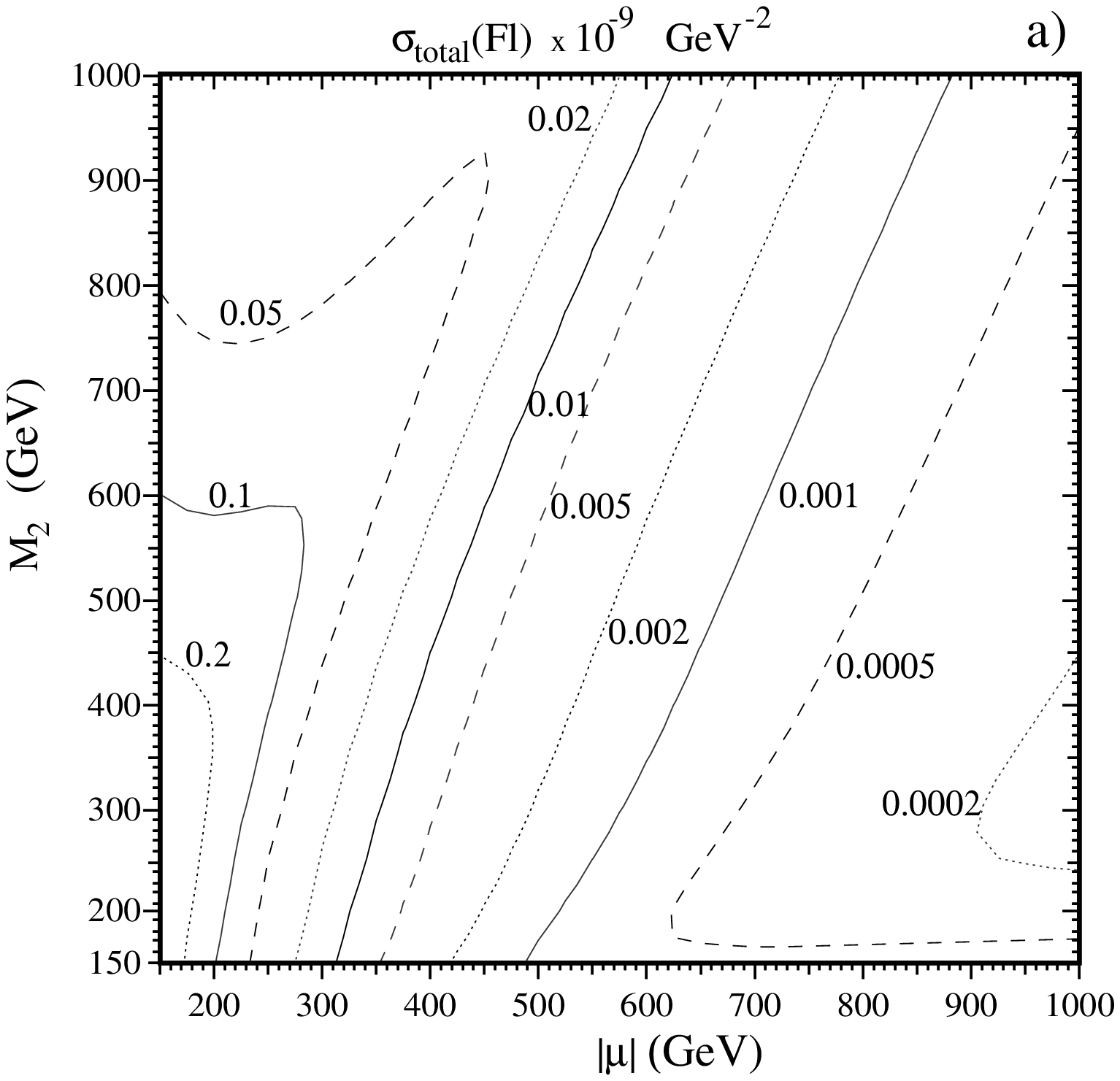,height=5.5in} 
 \hspace*{-0.8in}
\epsfig{file=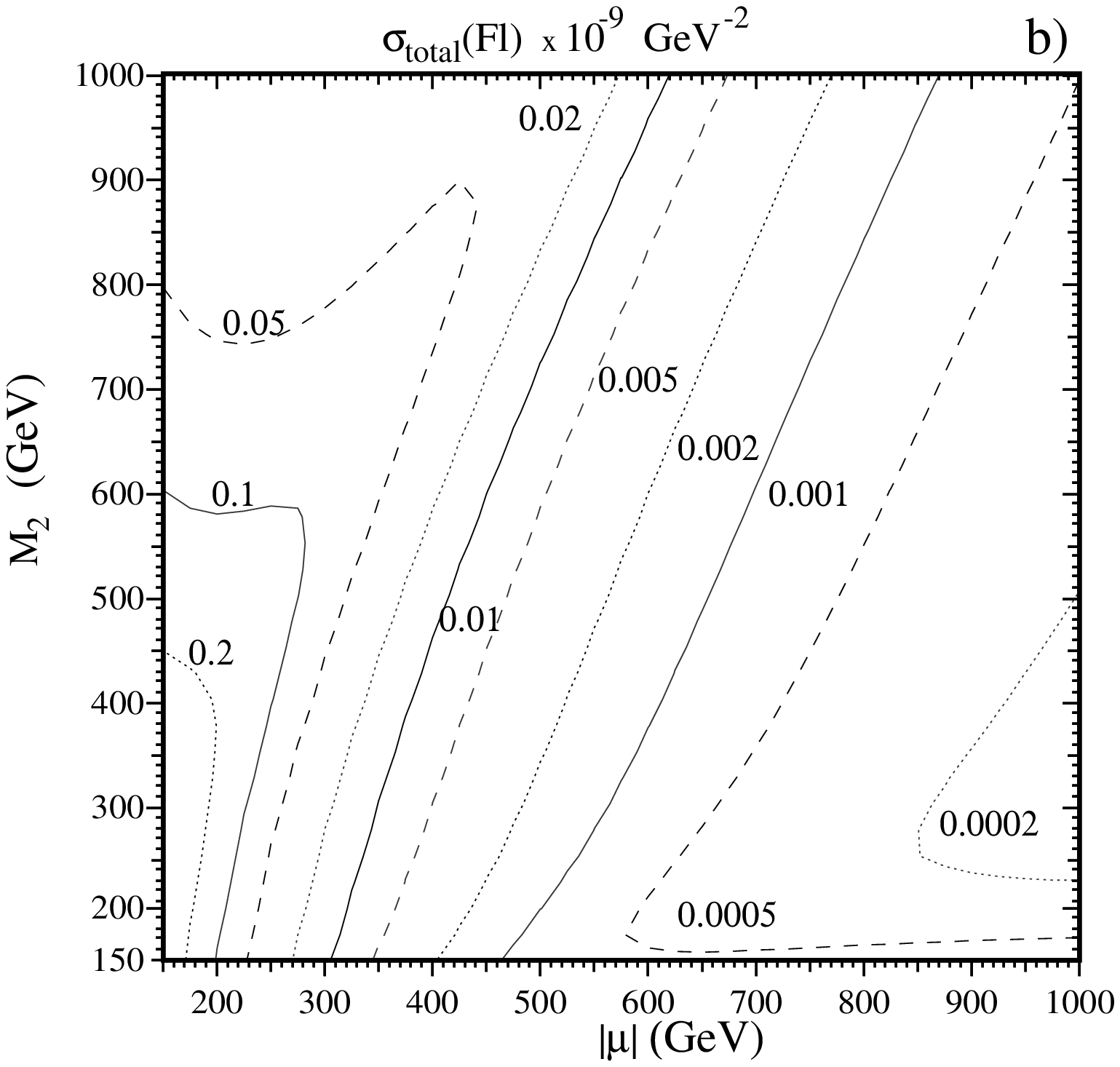,height=5.5in} 
\end{minipage}
\begin{minipage}{6.5in}
\vspace*{-1.9in}
\hspace{1.0in}
\epsfig{file=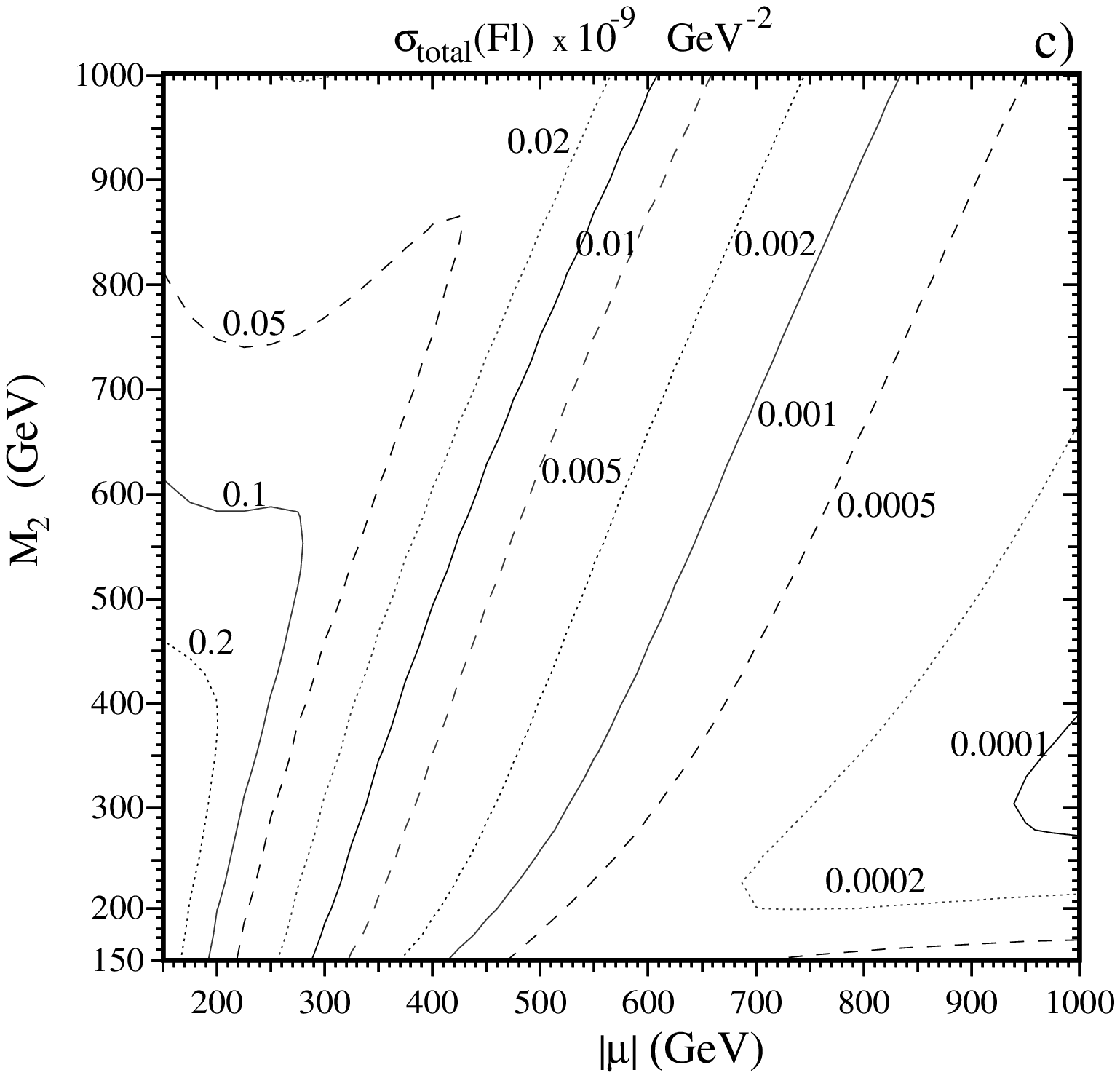,height=5.5in}
\end{minipage}
\vspace*{-0.6in}
\caption{\it The total cross-section for elastic scattering off of
$^{19}F$ a.)  for
$\theta_\mu = 
  0$, b) $\theta_\mu = \pi/8$, c.) $\theta_\mu = \pi/4$.  All plots have $\tha = \pi/2$.  The
  axis are in units of GeV.  The contours are in units of $10^{-9}$
  GeV$^{-2}$. }
\end{figure}

\begin{figure}
\vspace*{-1.5in} 
\hspace*{-0.5in}
\begin{minipage}{7.5in}
\epsfig{file=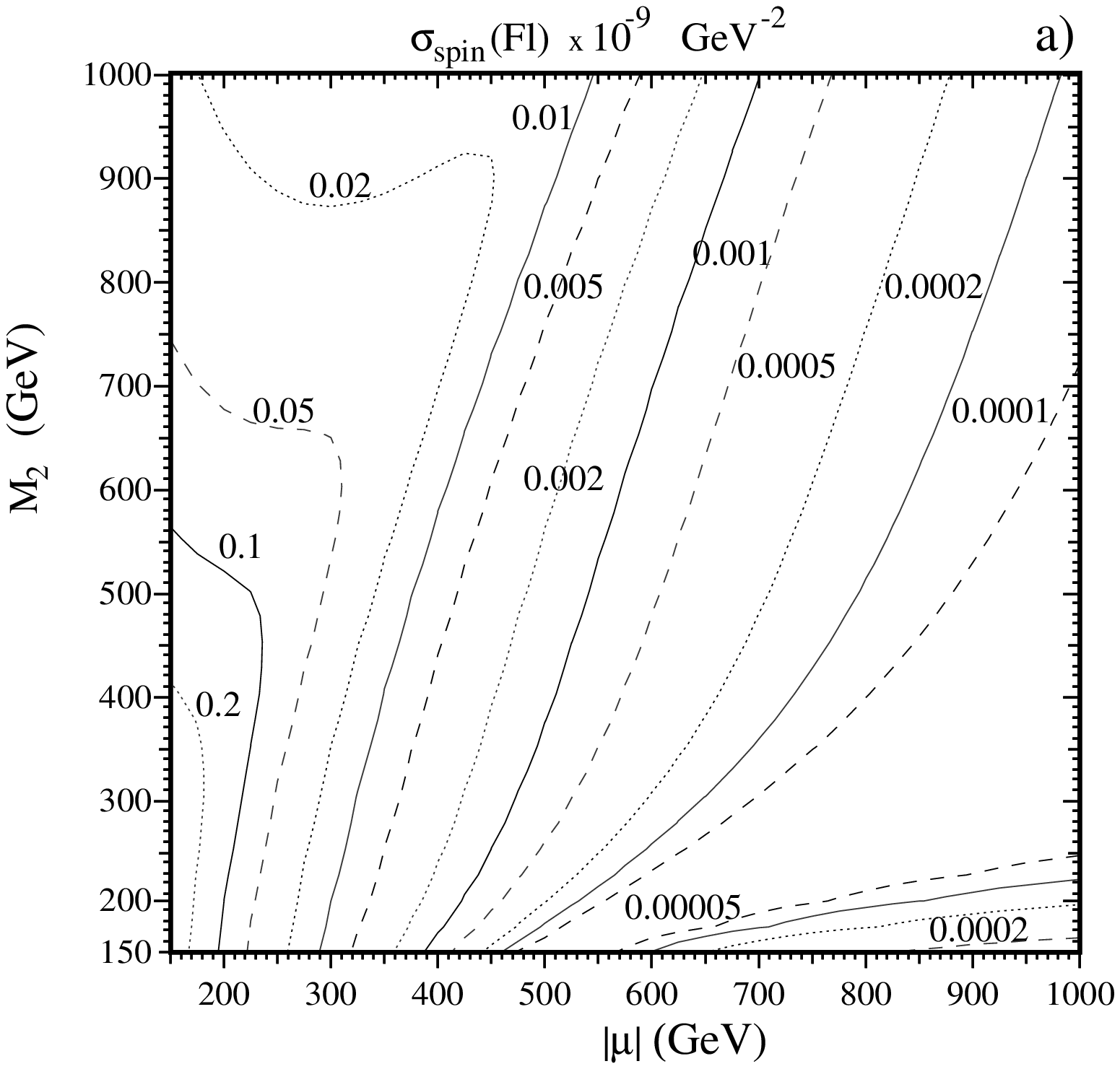,height=5.5in} 
 \hspace*{-0.8in}
\epsfig{file=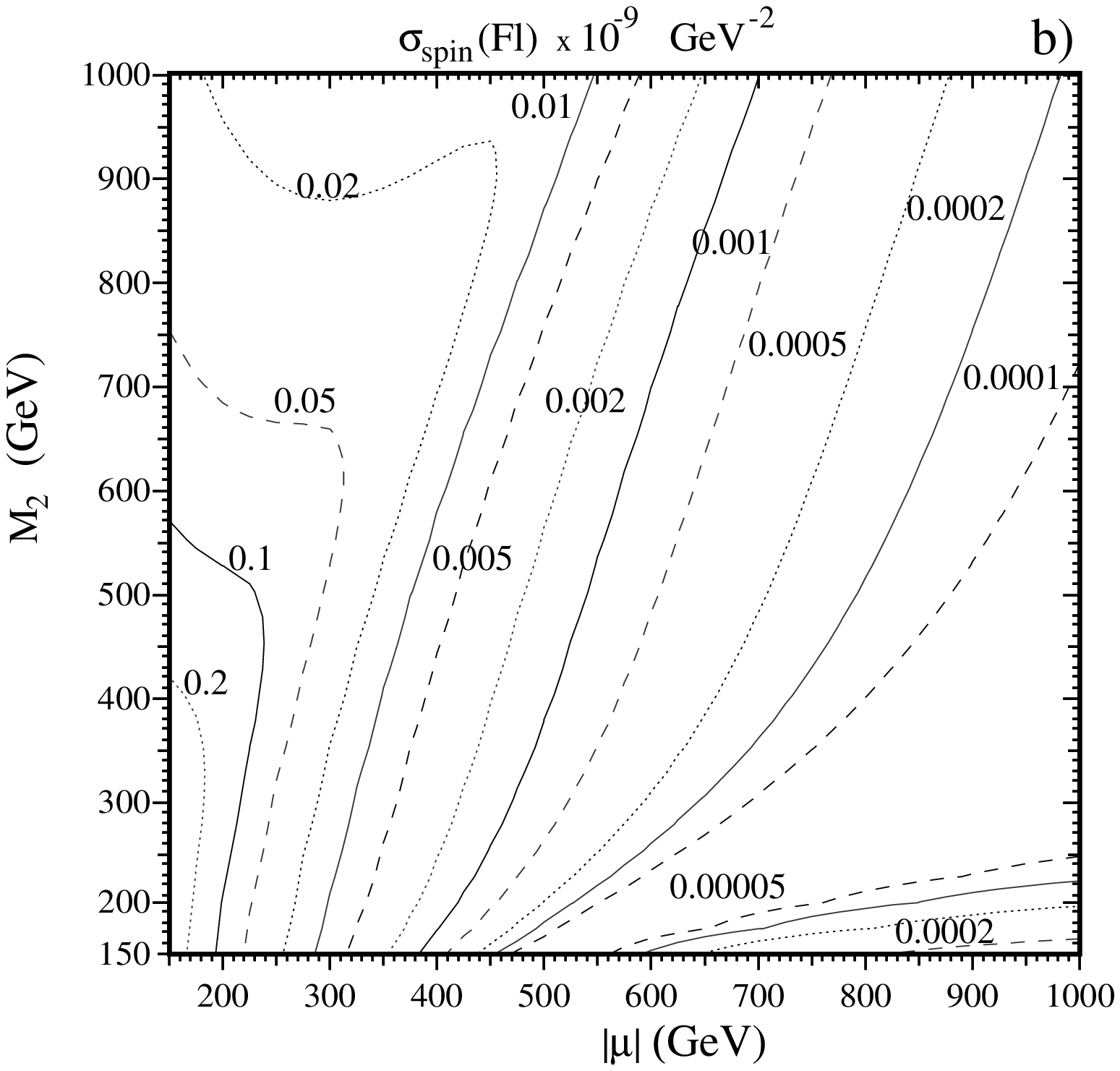,height=5.5in} 
\end{minipage}
\begin{minipage}{6.5in}
\vspace*{-1.9in}
\hspace{1.0in}
\epsfig{file=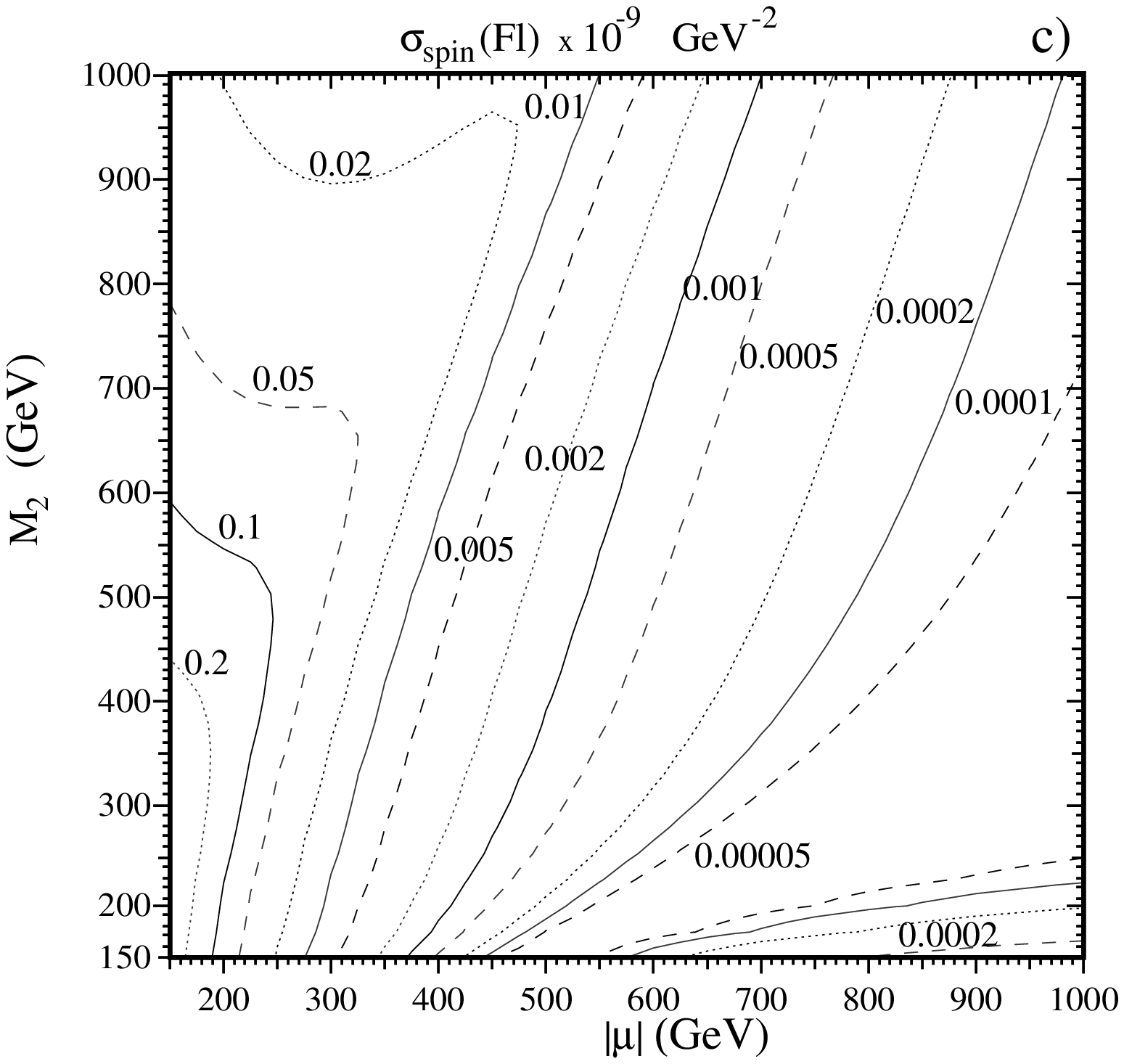,height=5.5in}
\end{minipage}
\vspace*{-0.6in}
\caption{\it The spin-dependent cross-section for elastic scattering
  off of $^{19}$F, for the same parameters as in Fig.~4.   The contours are in units of
$10^{-9}$ GeV$^{-2}$.}
\end{figure}

\begin{figure}
\vspace*{-1.5in} 
\hspace*{-0.5in}
\begin{minipage}{7.5in}
\epsfig{file=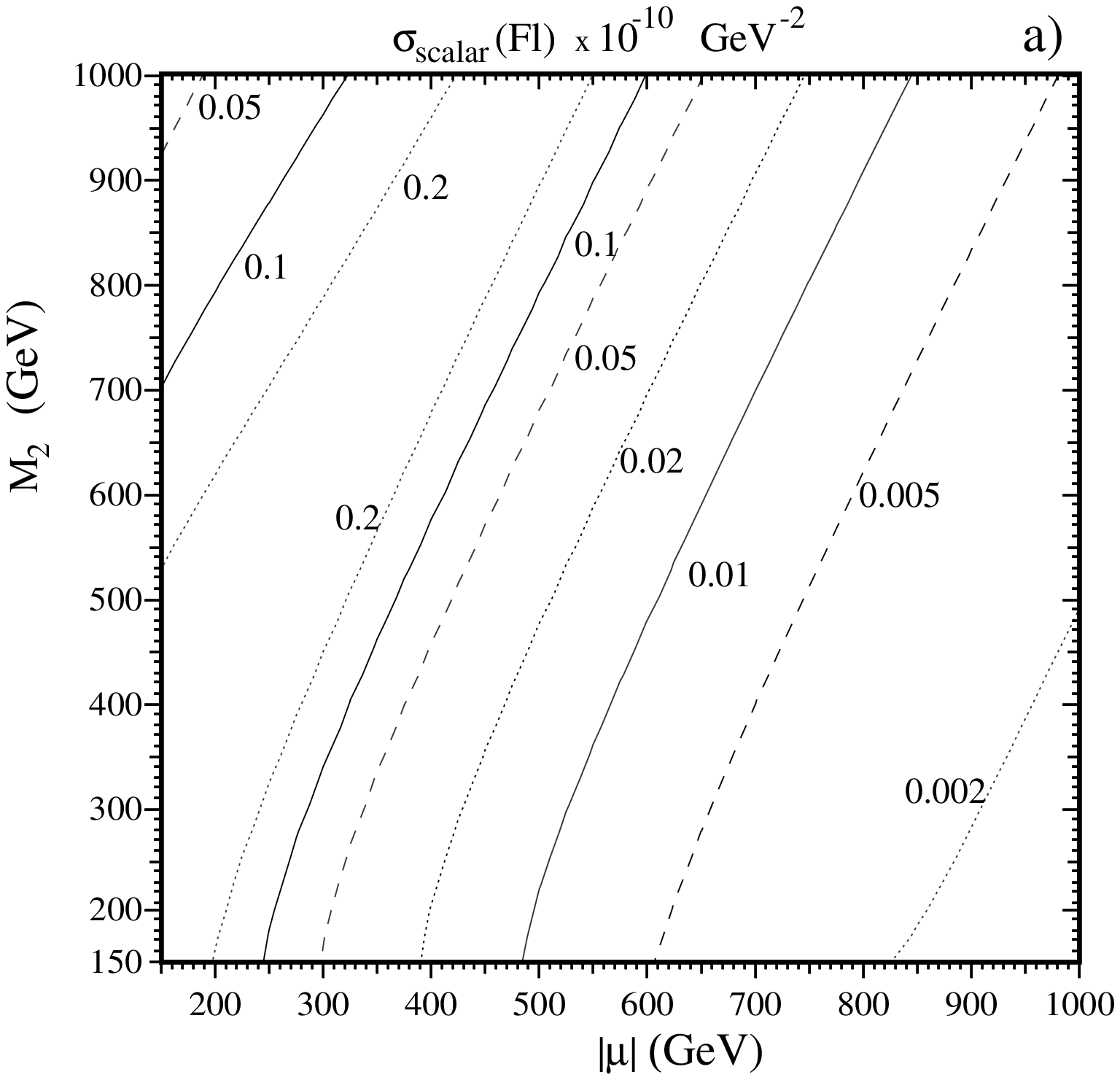,height=5.5in} 
 \hspace*{-0.8in}
\epsfig{file=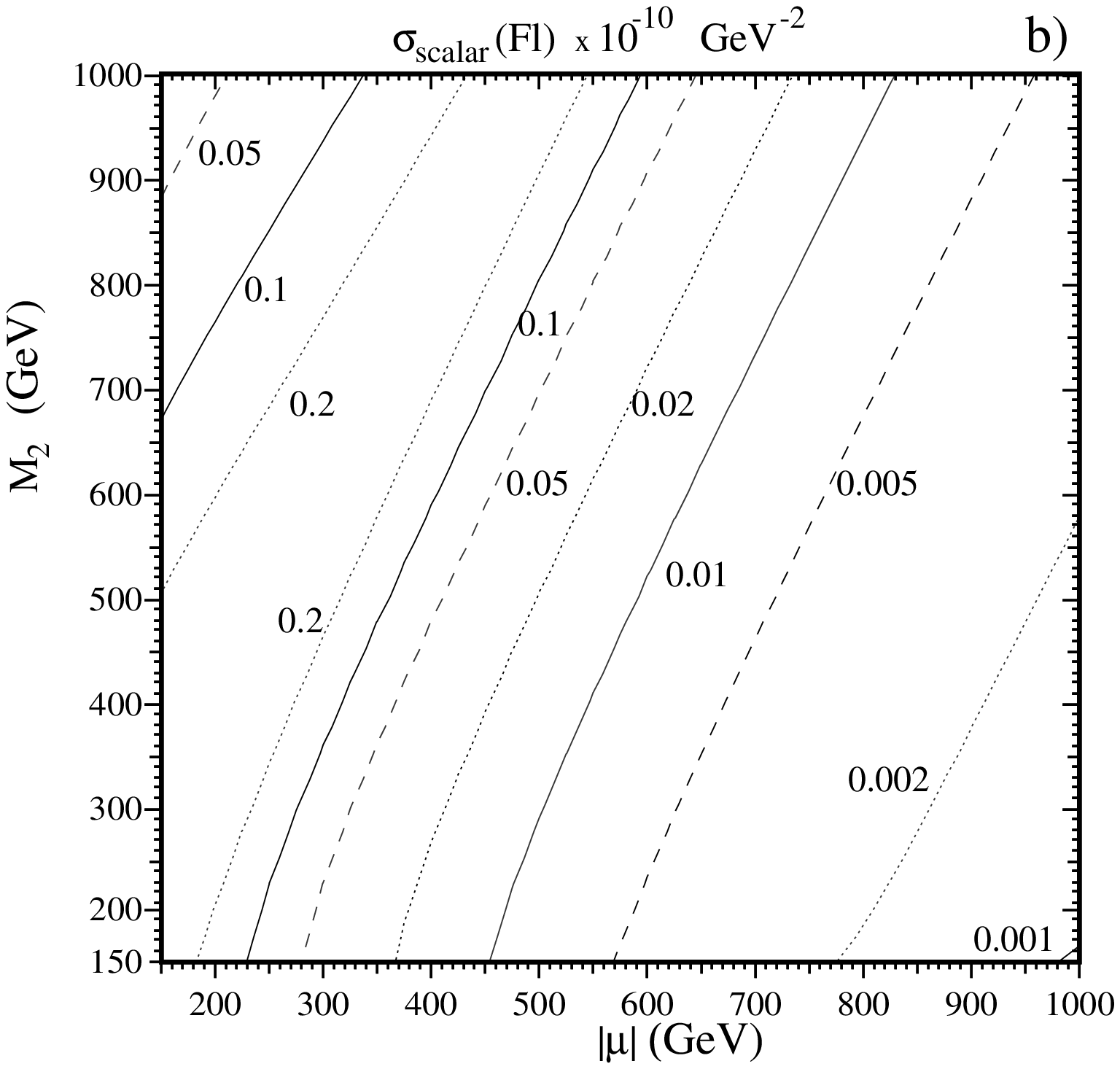,height=5.5in} 
\end{minipage}
\begin{minipage}{6.5in}
\vspace*{-1.9in}
\hspace{1.0in}
\epsfig{file=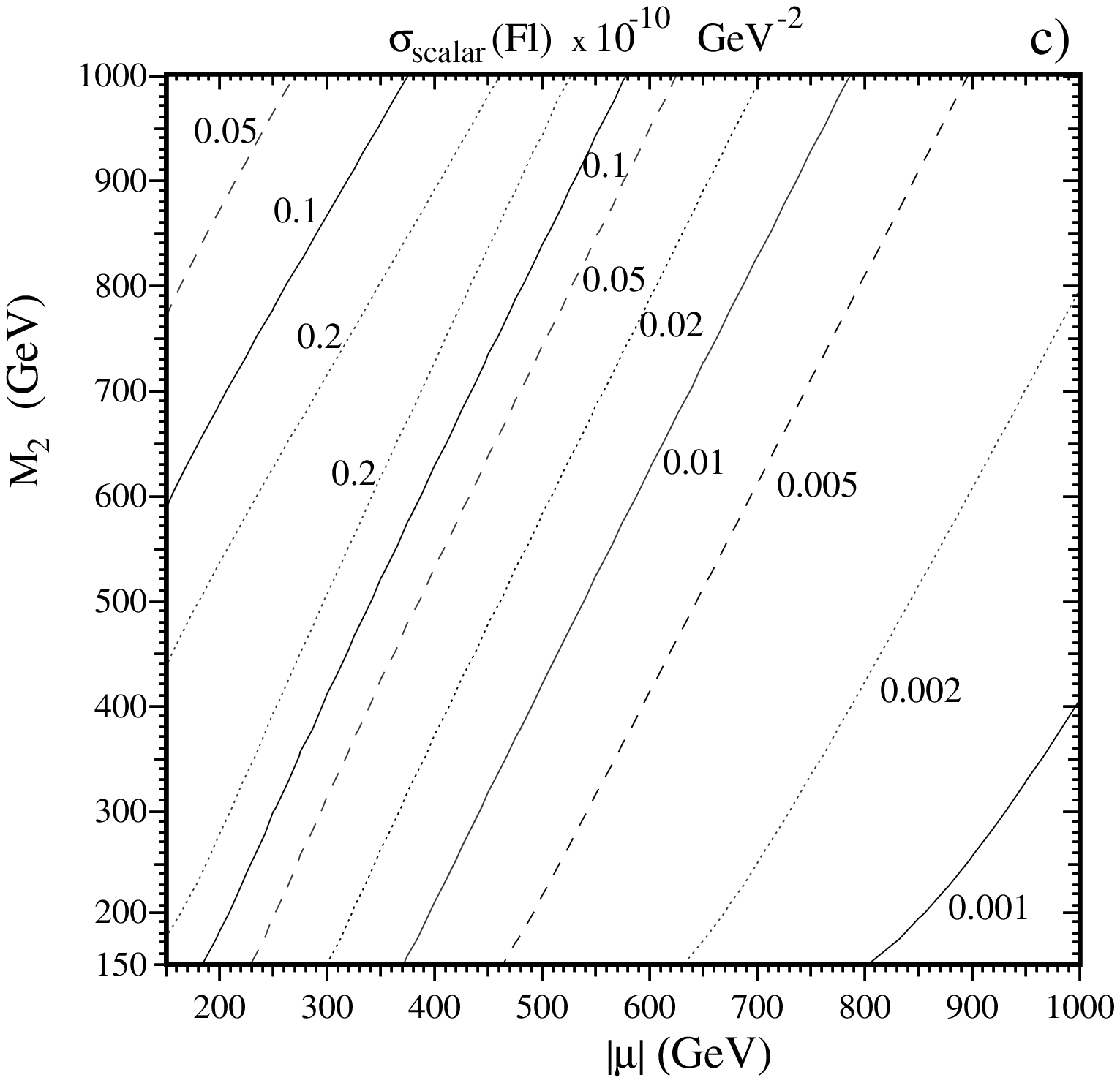,height=5.5in}
\end{minipage}
\vspace*{-0.6in}
\caption{\it  The spin-independent cross-section for elastic scattering
off of
$^{19}$F, for the same parameters as in Fig.~4. The contours are in units of $10^{-10}$
GeV$^{-2}$. }
\end{figure}

\begin{figure}
\vspace*{-2.5in}
\begin{center}\epsfig{file=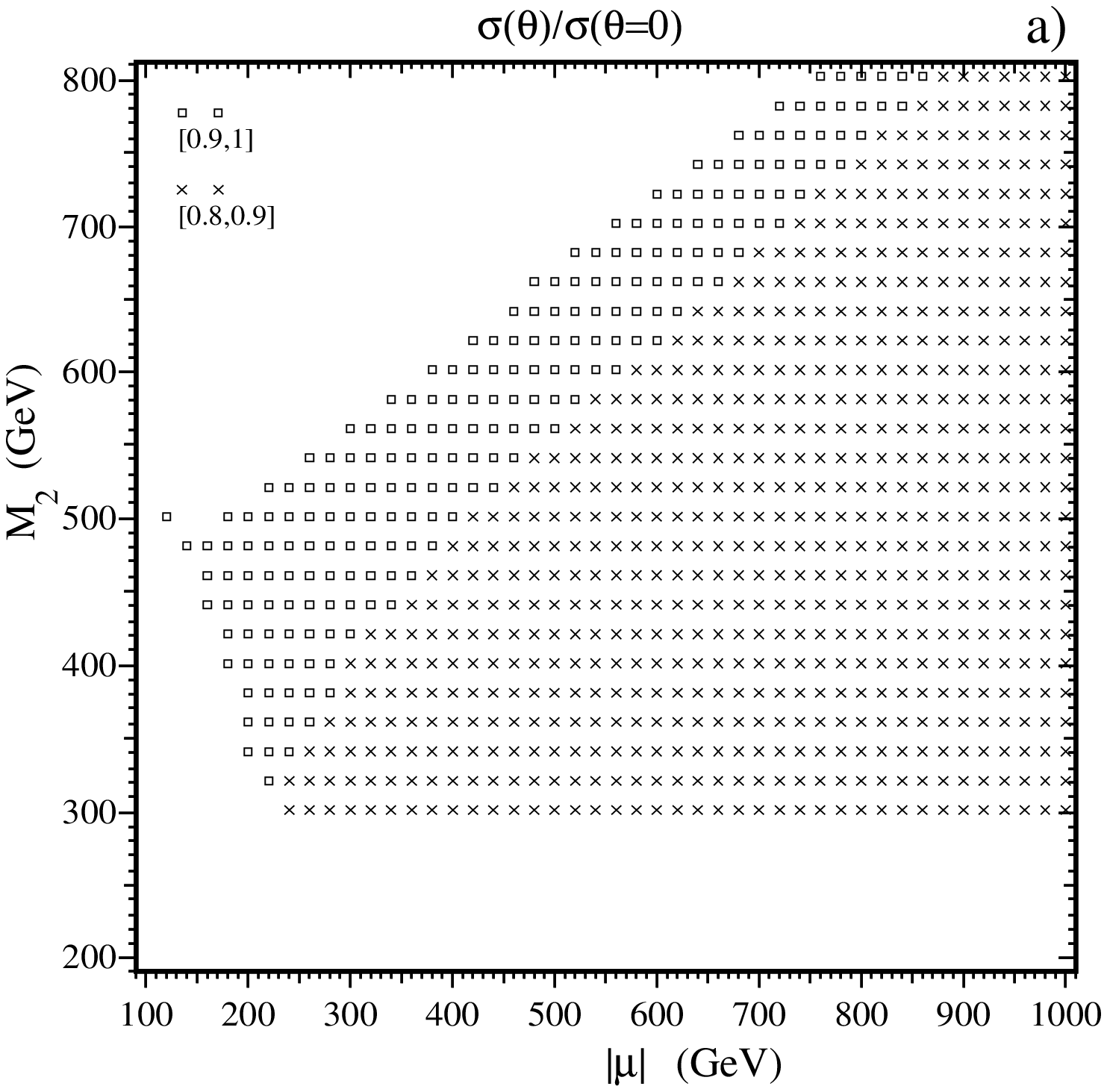,height=6.0in}\end{center}
\vspace*{-2.5in}
\begin{center}\epsfig{file=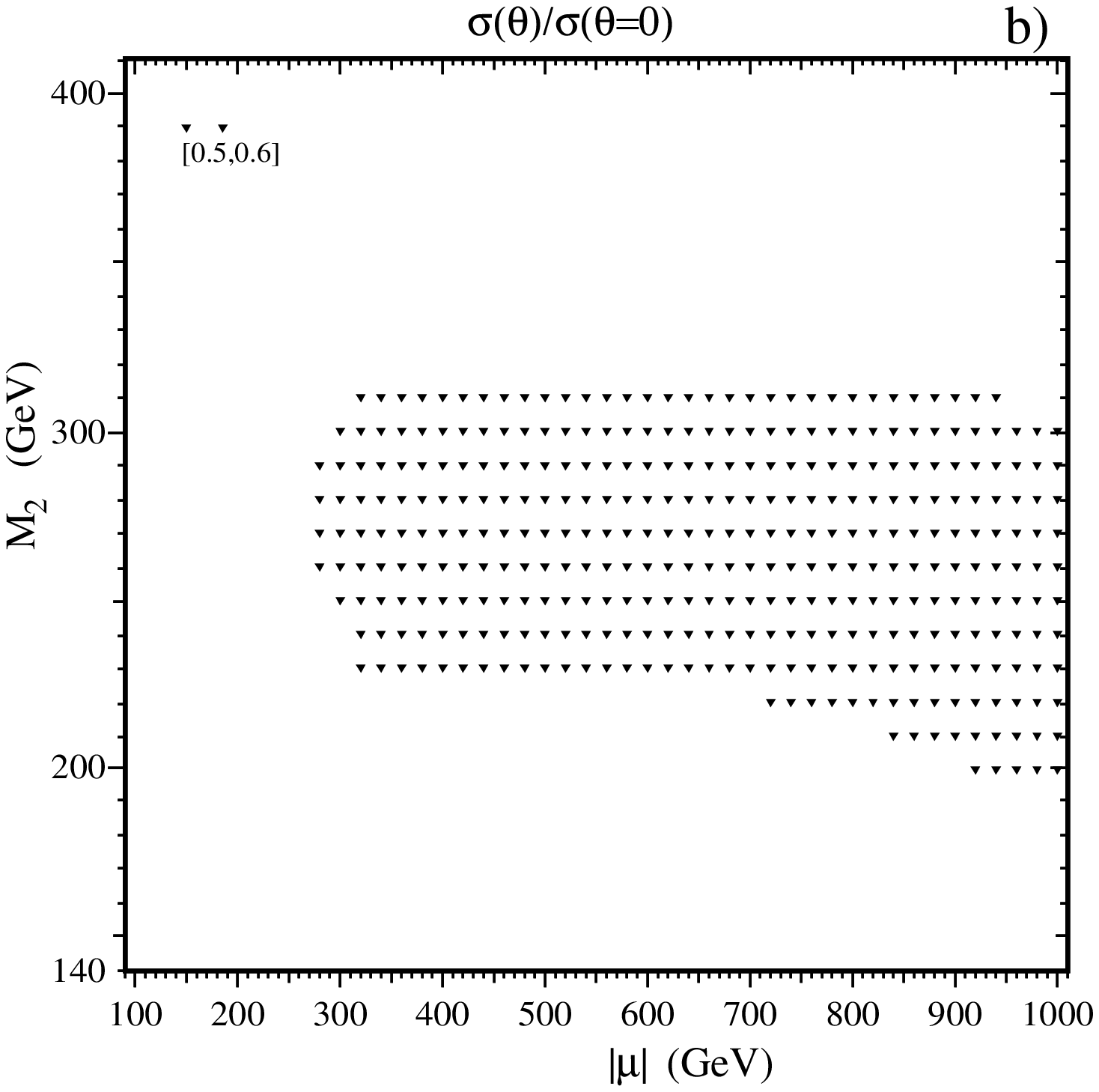,height=6.0in}\end{center}
\vspace*{-1.0in}
\caption{\it The ratio of the elastic scattering cross-section with
non-vanishing CP violating phases to the cross-section when the CP
violating phases are zero, for
$^{73}$Ge.  Here, the constraints from the eEDM and nEDM are taken into account
and the scan of the parameter space described in the text is projected
onto the $M_{2} - \mu$ plane.  In a) $\theta_\mu = \pi/8, \tha =
3\pi/8$  and b) $\theta_\mu = \pi/4$, $\tha =\pi/2$. }
\end{figure}

\begin{figure}
\vspace*{-2.5in}
\begin{center}\epsfig{file=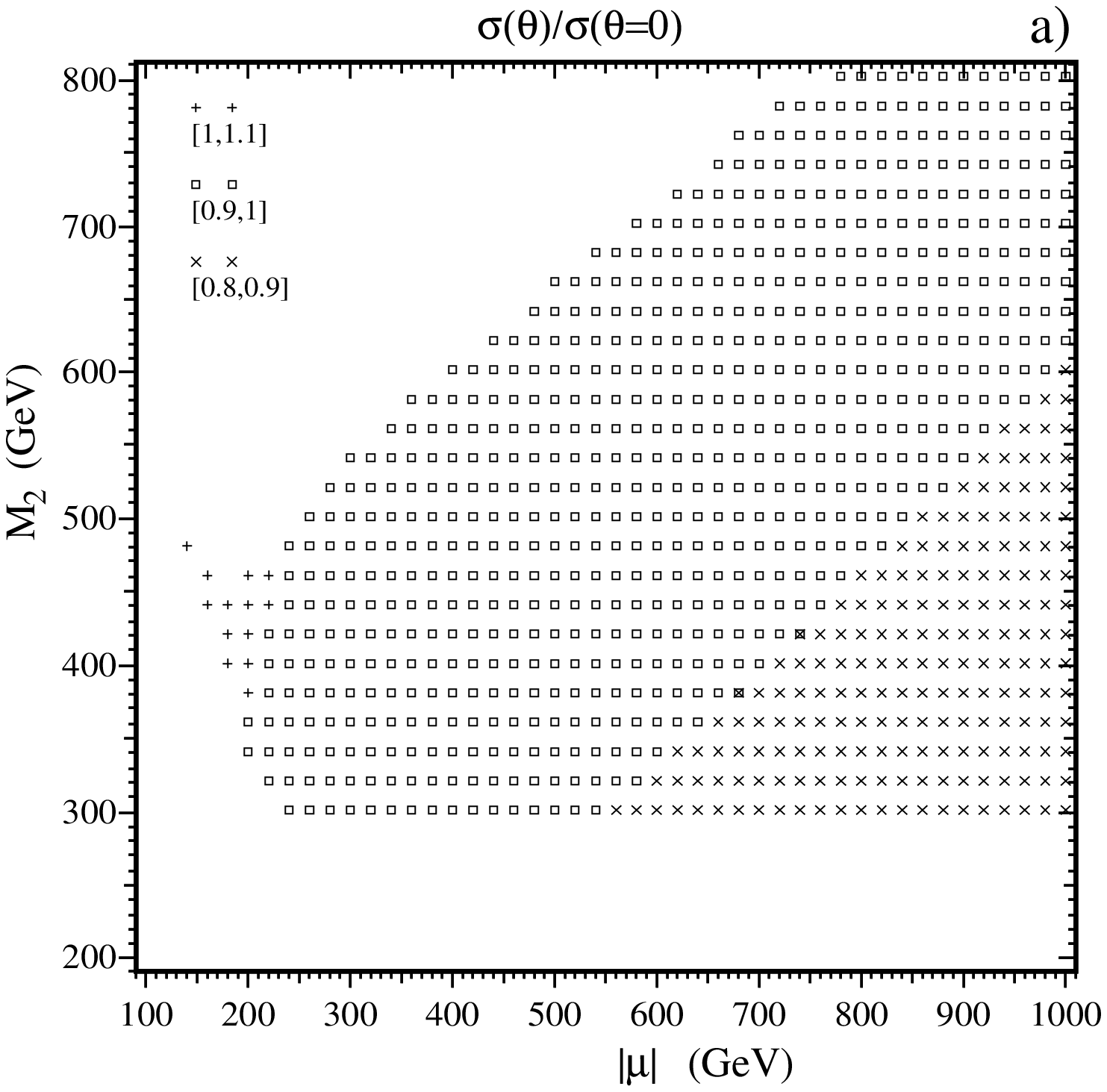,height=6.0in}\end{center}
\vspace*{-2.5in}
\begin{center}\epsfig{file=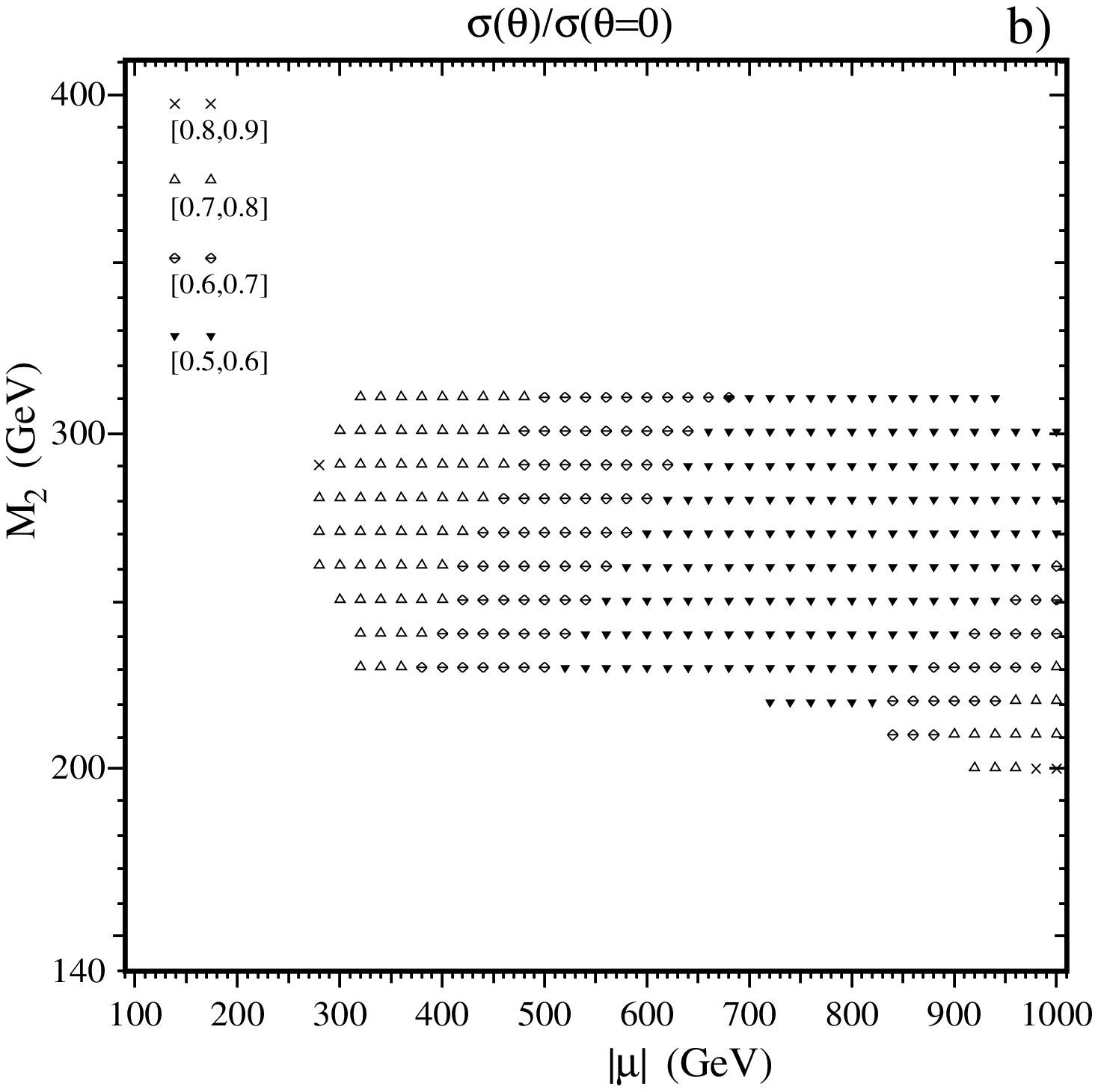,height=6.0in}\end{center}
\vspace*{-1.0in}
\caption{\it Same as Fig. 7 for $^{19}$F.}
\end{figure}

\begin{figure}
\vspace*{-2.5in}
\begin{center}\epsfig{file=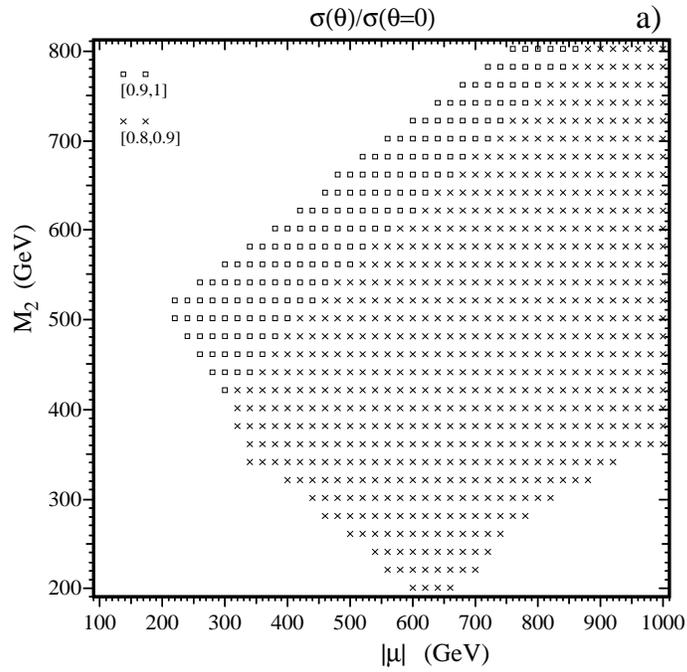,height=6.0in}\end{center}
\vspace*{-2.5in}
\begin{center}\epsfig{file=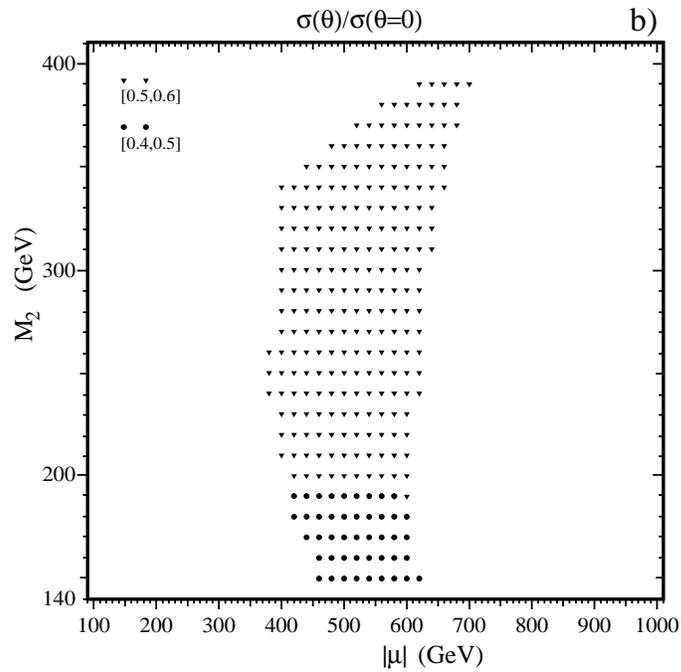,height=6.0in}\end{center}
\vspace*{-1.0in}
\caption{\it Same as Fig. 7 using the constraints from the eEDM and the HgEDM.}
\end{figure}

\begin{figure}
\vspace*{-2.5in}
\begin{center}\epsfig{file=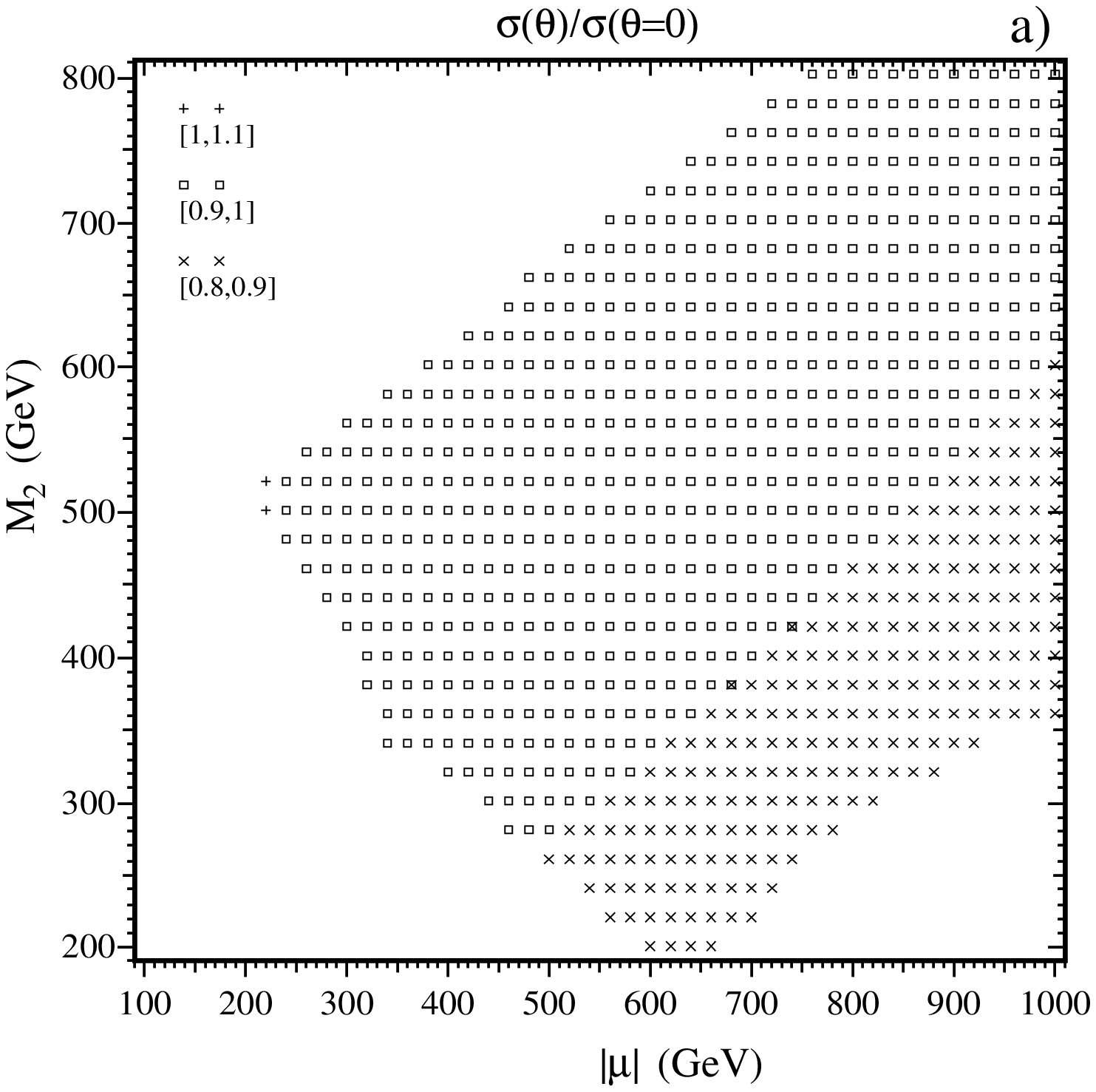,height=6.0in}\end{center}
\vspace*{-2.5in}
\begin{center}\epsfig{file=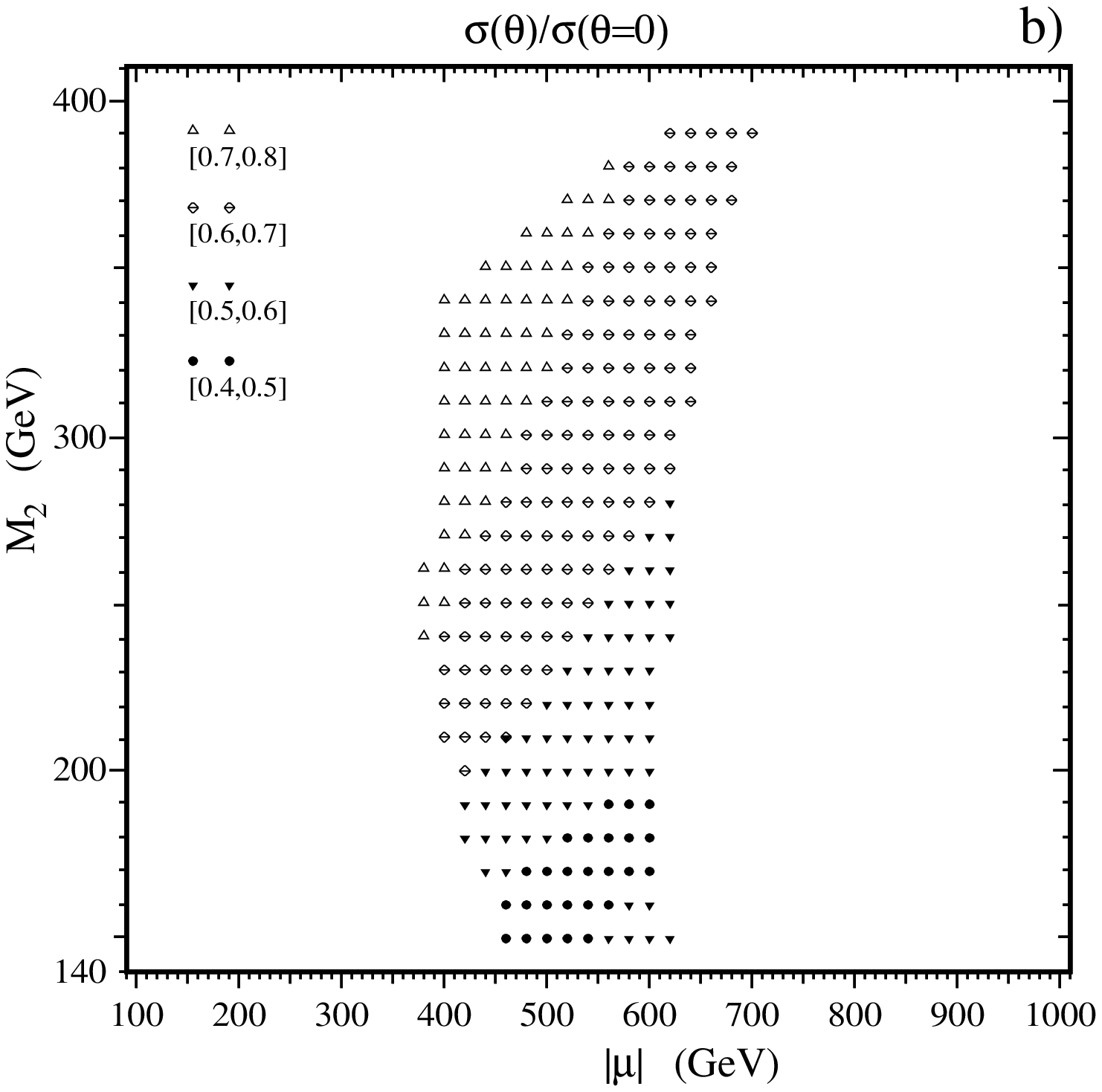,height=6.0in}\end{center}
\vspace*{-1.0in}
\caption{\it Same as Fig. 9 for $^{19}$F.}
\end{figure}


\vskip 1cm



\begin{thebibliography}{99}

\bibitem{MSSM} H.E. Haber and G.L. Kane, Phys. Rep. {\bf 117}
(1985) 75.


\bibitem{ehnos} J. Ellis, J.S. Hagelin, D.V. Nanopoulos, K.A. Olive
and M. Srednicki, Nucl. Phys. {\bf B238} (1984) 453.

\bibitem{suppl}
See, e.g., David O.\ Caldwell, in Proceedings of the Fifth 
International Workshop on Topics in Astroparticle and 
Underground Physics, 1997.

\bibitem{review}
 G.\ Jungman, M.\ Kamionkowski,
and K.\ Griest, \PR {\bf 267}, 195 (1996).

\bibitem{FFO}
T.\ Falk, A.\ Ferstl, and K.A.\ Olive, Phys. Rev. {\bf D59} (1999) 055009.

\bibitem{CIN}
U.\ Chattopadhyay, T.\ Ibrahim, and P.\ Nath,  hep-ph/9811362.

\bibitem{KS} S. Khalil and Q. Shafi, hep-ph/9904448.

\bibitem{dgh} M. Dugan, B. Grinstein and L. Hall, Nucl. Phys. {\bf B255}, 413
(1985).

\bibitem{smallthmu}
J.\ Ellis, S.\ Ferrara, and D.\ V.\ Nanopoulos, \PLB {\bf 114}
(1982) 231;  W.\ Buchm\"{u}ller and D.\ Wyler, \PLB {\bf 121} (1983)
321; J.\ Polchinski and M.\ Wise, \PLB {\bf 125} (1983) 393; 
F.\ del Aguila, M.\ Gavela, J.\ Grifols, and A.\ Mendez, \PLB 
{\bf 126} (1983) 71; C.\ V.\ Nanopoulos and M.\ Srednicki, \PLB
{ \bf 128 } (1983) 61. 

\bibitem{nath}P.\ Nath, Phys. Rev. Lett. {\bf 66} (1991), 2565.

\bibitem{KO} 
Y.\ Kizukuri and N.\ Oshimo, \PRD {\bf 45} (1992) 1806; 
\PRD {\bf 46} (1992) 3025.

\bibitem{FOS}
T.\ Falk, K.A.\ Olive, M.\ Srednicki  \PLB {\bf 354}, 99 (1995).

\bibitem{heavy}S.~Dimopoulos and G.F.~Giudice, Phys.~Lett.~{\bf B357}
(1995)
   573 ; A.~Cohen, D.B.~Kaplan and A.E.~Nelson,  Phys.~Lett.~{\bf B388}
(1996)
   599 ; A.~Pomarol and D.~Tommasini, Nucl.~Phys.~{\bf B488} (1996)  3.

\bibitem{fko1}  T. Falk and K.A. Olive, Phys. Lett. {\bf B375} (1996) 196. 

\bibitem{IN2}
T.\ Ibrahim, P.\ Nath,  \PRD {\bf 57} (1998) 478;  
Erratum \PRD {\bf 58}, 019901 (1998); T.\ Ibrahim, P.\ Nath, \PLB {\bf
  418} (1998) 98; T.\ Ibrahim, P.\ Nath,  \PRD {\bf 58} (1998) 111301.

\bibitem{fko2}  T. Falk and K.A. Olive, Phys. Lett. {\bf B439} (1998) 71. 

\bibitem{bgk}  T. Ibrahim and P. Nath, Phys. Rev. {\bf D58} (1998) 111301; 
M. Brhlik, G. J. Good and G.L. Kane, Phys. Rev. {\bf D59} (1999) 115004.

\bibitem{prs} S. Pokorski, J. Rosiek, and C.A. Savoy, hep-ph/9906206.


\bibitem{phm} A. Pilaftsis, Phys. Rev. D {\bf 58}  (1998) 096010;
  A. Pilaftsis, Phys.  Lett. {\bf 435B} (1998) 88; D. A. Demir, hep-ph/9901389; D. A. Demir,
  hep-ph/9905571.

\bibitem{pw}A. Pilaftsis and C.E.M. Wagner, hep-ph/9902371.

\bibitem{COPW}
M.\ Carena, M.\ Olechowski, S.\ Pokorski, and C.\ E.\ M.\ Wagner,
\NPB {\bf 426}, 269 (1994).

\bibitem{ef}
J.\ Ellis and R.\ Flores, \PLB {\bf 300}, 175 (1993).


\bibitem{formfactor}
H.\ Cheng, \PLB {\bf 219}, 347 (1989).

\bibitem{ef2}
J.\ Ellis and R.\ Flores, \NPB {\bf 307}, 883 (1988).


\bibitem{griest}
K.\ Griest \PRD {\bf 38}, 2357 (1988).

\bibitem{adams}
The Spin Muon Collaboration, D.\ Adams {\it et.\ al.}, \PLB 
{\bf 329}, 399 (1994).

\bibitem{SVZ}
M.\ A.\ Shifman, A.\ I.\ Vainshtein, and V.\ I.\ Zakharov, 
{\em Phys.\ Lett.} {\bf 78B}, 443 (1978); \\ A.\ I.\ Vainshtein,
V.\ I.\ Zakharov, M.\ A.\ Shifman, {\em Usp.\ Fiz.\ Nauk} 
{\bf 130}, 537 (1980).

\bibitem{GLS}
J.\ Gasser, H.\ Leutwyler, and M.\ E.\ Sainio, \PLB {\bf 253}, 
252 (1991).

\bibitem{DN}
M. Drees and M.\ M.\ Nojiri, \PRD {\bf 47}, 4226 (1993);
M. Drees and M.\ M.\ Nojiri, \PRD {\bf 48}, 3483 (1993).

\bibitem{nedm} I. S. Altarev {\it et al.}, Phys. Lett {\bf B276} (1992) 242.

\bibitem{eedm} E. Commins {\it et al.}, Phys. Rev. {\bf A50} (1994) 2960.

\bibitem{medm} J.P. Jacobs {\em et al.}, Phys. Rev. Lett. {\bf 71} (1993) 
3782; J.P. Jacobs {\em et al.}, Phys. Rev. {\bf A52} (1995) 3521.


\bibitem{wads} S. Weinberg, Phys. Rev. Lett. {\bf 63} (1989) 2333;
R. Arnowitt, M. Duff, and K. Stelle, Phys. Rev. {\bf D43} (1991) 3085.
 

\bibitem{fopr} T. Falk, K.A. Olive, M. Pospelov, and R. Roiban, hep-ph/9904393.

\bibitem{krip} V.M. Khatsimovsky, I.B. Khriplovich and A.R. Zhitnitsky, Z. 
Phys. {\bf C36} (1987) 455; V.M. Khatsimovsky, I.B. Khriplovich and A.S. Yelkhovsky,
Ann.  Phys. {\bf 186} (1988) 1;  V.M. Khatsimovsky and I.B. Khriplovich, Phys. Lett.
{\bf B296}  (1994) 219. 

\bibitem{garisto}
R.\ Garisto and J.\ Wells \PRD {\bf 55}, 1611 (1997).

\bibitem{aad}E. Accomando, R. Arnowitt and B. Dutta, hep-ph/9907446.

\bibitem{lep} Official compilations of LEP limits on supersymmetric
particles are available from:\\
{\tt http://www.cern.ch/LEPSUSY/}





\bibitem{efos} J.~Ellis, T.~Falk, K.A.~Olive and M.~Schmitt,
Phys. Lett. {\bf B413} (1997) 355; J.\ Ellis, T.\ Falk, G.\ Ganis, K.A.\ Olive, M.\
Schmitt,
\PRD {\bf 58} (1998) 095002; J. Ellis, T. Falk, and K.A. Olive, Phys. Lett. {\bf B444}
(1998) 367; J. Ellis, T. Falk, K.A. Olive, and M. Srednicki, hep-ph/9905481.
 



\end{thebibliography}
\end{document}